\documentclass[journal,twocolumn]{IEEEtran}
\usepackage{graphicx}
\usepackage{cite}
\usepackage{amsmath,amsfonts}
\usepackage{times}
\usepackage{amssymb}
\usepackage{stfloats}
\usepackage{array}
\usepackage{textcomp}
\usepackage{url}
\usepackage{color}
\usepackage{multirow}
\usepackage{algorithmic}
\usepackage{verbatim}
\usepackage{newclude}
\usepackage{setspace}
\usepackage{fancyhdr}
\usepackage{subcaption}
\usepackage{epstopdf}
\usepackage{algorithm}

\ifCLASSINFOpdf

\else

\fi

\begin{document}

\title{Wideband Beamforming with RIS: A Unified Framework via Space-Frequency Transformation}

\author{\IEEEauthorblockN{Xiaowei Qian, Xiaoling Hu, {\em Member, IEEE},   Chenxi Liu, {\em Senior Member, IEEE} and Mugen Peng, {\em Fellow, IEEE}}

\thanks{Xiaowei Qian, Xiaoling Hu, Chenxi Liu and Mugen Peng are with the State Key Laboratory of Networking and Switching Technology, Beijing University of Posts and Telecommunications, Beijing 100876, China (e-mail: \{xiaowei.qian, xiaolinghu, chenxi.liu, pmg\}@bupt.edu.cn).}}

% The paper headers

\maketitle

\begin{abstract}
\color{black}
The spectrum shift from the sub-6G band to the high-frequency band has posed an ever-increasing demand on the paradigm shift from narrowband beamforming to wideband beamforming. Despite recent research efforts, the problem of wideband beamforming design is particularly challenging in reconfigurable intelligent surface (RIS)-assisted systems, due to that RIS is not capable of performing frequency-dependent phase shift, therefore inducing high signal processing complexity.
In this paper, we propose a simple-yet-efficient wideband beamforming design for RIS-assisted systems, in which a transmitter sends wideband signals to a desired target, through the aid of the RIS.
In our proposed design, we exploit space-frequency Fourier transformation and stationary phase method to yield an approximate closed-form solution of the RIS phase shifts which significantly reduces the signal processing complexity, compared to the existing approaches. The obtained solution is then used to generate a large and flat beampattern over the desired frequency band. 
Through numerical results, we validate the effectiveness of our proposed beamforming design and demonstrate how it can improve system performances in terms of communication rate and sensing resolution.
Beyond generating the flat beampattern, we highlight that our proposed design is capable of mimicking any desired beampattern by matching the RIS phase shift with the amplitude modulation function, thus providing valuable insights into the design of novel wideband beamforming for RIS-assisted systems.

\end{abstract}


\IEEEpeerreviewmaketitle

\section{Introduction}
\label{SEC_introduction}
The reconfigurable intelligent surface (RIS) has gained significant attention in recent years and is regarded as a promising technology for the sixth-generation (6G) networks \cite{gong2020toward}. The RIS is a programmable electromagnetic (EM) metasurface with the structure of a two-dimensional (2D) planar array composed of a large amount of low-cost passive reflecting elements, which can manipulate the EM properties of the incident signals \cite{cui2014coding}. Particularly, by adjusting the phase of the incident signals, the RIS can perform beamforming passively without any radio frequency (RF) hardware and additional energy consumption \cite{zeng2020reconfigurable}, which is especially attractive for high-frequency wireless systems, e.g., the millimeter-wave (mmWave) and terahertz (THz) system, compared with the traditional active phased-array antenna (PAA).

Since high-frequency signals suffer from considerable attenuation and are susceptible to obstruction \cite{chenwenyun_2024_UAV_TWC}, the employment of RIS beamforming has been widely investigated for boosting signal strengths in a low-cost manner \cite{wu2021intelligent}. Specifically, RIS beamforming can steer the signals toward the desired direction/position to effectively enhance the signal-to-noise ratio (SNR), thereby improving both the communication performance \cite{wu2019intelligent,RIS_ISAC_Wang_2022,yuzhouyuan_2022_beamforming,Qian2023_myself_TCOM} and sensing performance \cite{Lilianlin_nature_2019MachinelearningRM,moro2024exploring_UAV_ISAC_imaging,RIS_BF_ISAC_2022_JSAC,RIS_localization_2022}. 
For example, in \cite{wu2019intelligent}, the authors study a RIS-assisted wireless system, minimize the total transmit power at the transmitter through the joint optimization of transmit beamforming at the transmitter and passive beamforming at the RIS, and prove that the transmit power decreases with the increase of the number of RIS reflecting elements $N_{\mathrm{RIS}}$ in the order of $N_{\mathrm{RIS}}^2$ when $N_{\mathrm{RIS}}$ is sufficiently large. Additionally, in \cite{RIS_localization_2022}, authors propose a RIS-aided low-complexity codebook-based method for joint localization and synchronization. The numerical analysis shows that the proposed joint beamforming scheme provides enhanced localization performance.

Although numerous studies have demonstrated that RIS beamforming enhances communication and sensing performance, their scope has been limited to narrowband systems. However, with the escalating demands for higher data rates and better sensing capabilities, the trend toward higher frequency and wider bandwidth has become prevalent \cite{liu2022integrated}. Unfortunately, the narrowband RIS beamforming is no longer applicable in wideband scenarios due to the beam squint effect \cite{cui2022near_rainbow_beamsquint}. The beam squint is the phenomenon where the beamforming direction changes with the operating frequency \cite{mailloux2017phased}, which will misdirect the beam, cause dramatic gain variation over the desired bandwidth, and distort the wideband waveform of the incident signal in the expected direction or position. As a result, it limits the effective bandwidth, reducing the wideband communication rate and deteriorating the distance resolution. 

In traditional active PAA systems, beam squints can be mitigated through the time-delay beamforming architectures, such as true time delay (TTD) array and delay-phase precoding structure \cite{cui2022near_rainbow_beamsquint,liu2010wideband_beamforming_book}. For example, a common approach to mitigate beam squint is to use TTD-based arrays where the delay of the RF signal can be electronically controlled at each antenna. However, the above beam squint mitigation approaches are no longer suitable for the RIS wideband beamforming because the RIS has no traditional TTD structures. 

{\color{black}
Several studies have investigated the RIS wideband beamforming techniques \cite{hao2022ultra_RIS_survey_THZ,hao2023far_near_beam_squint_IRS_recent_work_TVT,TDU_based_RIS_comm,SDR_WBF_RIS_2022,RIS_WBF_TIME_DELAY_at_BS_2023}. There are two approaches to solve the RIS wideband beamforming problem. One is to add time-delay units at the RIS \cite{hao2023far_near_beam_squint_IRS_recent_work_TVT,TDU_based_RIS_comm}. For example, in \cite{hao2023far_near_beam_squint_IRS_recent_work_TVT}, authors propose to apply delay adjustable units to the RIS and jointly optimize the phase shifts and time delays of RIS elements, which eliminates the gain loss caused by beam squint in both near-field and far-field scenarios. Similarly, in \cite{TDU_based_RIS_comm}, the authors further propose to group the adjacent RIS elements into a sub-surface to share a common time-delay unit for lowering the cost and energy consumption of the time-delay unit. 
Another one is to re-design the phase shifts via convex optimization \cite{SDR_WBF_RIS_2022,RIS_WBF_TIME_DELAY_at_BS_2023}. For example, in \cite{SDR_WBF_RIS_2022}, authors perform a majorization-minimization-based iterative approach to optimize the RIS passive beamforming such that the sum rate over all subcarriers is maximized. In \cite{RIS_WBF_TIME_DELAY_at_BS_2023}, the authors propose to jointly design hybrid analog/digital beamforming, time delays at the BS, and the phase shifts at the RIS via typical convex optimization methods like Lagrangian dual reformulation, multidimensional complex quadratic transform, and alternating direction method of multipliers.
}

For the first approach, the delay adjustable units are too complicated and expensive to integrate into the low-cost passive RIS at this stage. For the second approach, the RIS wideband beamforming problem is only formulated as an optimization problem and solved via traditional convex optimization methods, which leads to intolerable computational complexity when the size of the RIS is large. This raises the question: Is there a low-complexity and low-cost RIS wideband beamforming method that is adaptable even if the number of RIS elements is very large, e.g., more than $10^6$? 

Motivated by this, in this paper, we address the fundamental problem of RIS wideband beamforming. The objective is to generate a wideband beam pointing toward a desired target (DT), with a large and flat gain over the desired frequency band. Our main contributions are summarized as follows.

\begin{itemize}
    \item 
    Rather than following the traditional approaches of adding time-delay units or performing complicated convex optimization, we propose a novel methodology by exploiting space-frequency Fourier transformation (SFFT) and stationary phase method (SPM) to solve the RIS wideband beamforming problem for the first time. 

     \item 
    \textcolor{black}{Compared with the previous RIS wideband beamforming methods, our method does not require additional time-delay units and yields an approximate closed-form solution of RIS phase shifts, allowing the realization of a large and flat gain over the desired frequency band with extremely low complexity, making it highly adaptable when dealing with large RIS arrays in wideband cases.}
    
    \item 
    Simulation results demonstrate the advantages of the proposed RIS wideband beamforming over the traditional RIS narrowband beamforming in terms of both communication rate and sensing resolution. Specifically, by applying the proposed RIS wideband beamforming, the communication rate improves by over 10\%. More importantly, we present a first-time finding that the distance resolution in sensing can be significantly enhanced with the RIS wideband beamforming.
   
    \item 
    The idea of applying SFFT to the design of a flat beampattern can be generalized to the design of any expected frequency-domain beampattern by matching the RIS phase shift with the amplitude modulation function. This indicates its great potential in RIS-assisted wideband applications through establishing space and frequency relationships.
\end{itemize}

The remainder of this paper is organized as follows. Section \ref{sec:system_model} introduces the system model of the RIS wideband beamforming architecture. Section \ref{sec:RIS Wideband Beamforming Design} presents the SFFT-based RIS wideband beamforming design. Numerical results are provided in Section \ref{sec:simulation results}. Finally, Section \ref{sec:conclusion} concludes this paper.

\section{System Model}
\label{sec:system_model}
\subsection{System Architecture}
\label{subsec_system_model_A}
\begin{figure}[t]
    \centering
    \includegraphics[width=0.6\linewidth]{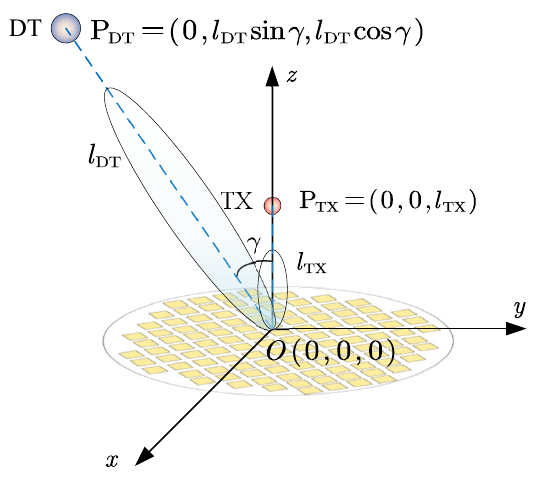}
    \caption{System model.}
    \label{model_illustration}
\end{figure}
We consider the near-field RIS wideband beamforming problem, whose objective is to generate a wideband beam pointing toward a DT, such as a communication user, a sensing target, or an ISAC user. As shown in Fig.\ref{model_illustration}, a TX transmits the wideband signals with the bandwidth $B$ and the carrier frequency $f_{\text{c}}$, while a RIS reflects and manipulates the transmitted signals from the TX to the DT. 

We first establish the Cartesian coordinate system $\mathcal{C}_0$ where the center of the RIS is defined as the origin $O$ with coordinates $(0,0,0)$. The RIS's boresight, i.e., the normal direction passing through the center of the RIS, is defined as the $z$-axis. For ease of analysis, we consider that the RIS is circular and positioned on the $x-O-y$ plane. Owing to the inherent symmetry, the $x$-axis and $y$-axis can be selected as any mutually perpendicular directions in the plane of the RIS panel, both originating from the origin point $O$. 
Hence, for simplicity, we place the TX along the positive half of the $z$-axis, and the DT on the $y-O-z$ plane. Specifically, the TX is located at the $\mathrm{P}_{\mathrm{TX}}$, with coordinates $(0,0,l_{\text{TX}})$, where $l_{\text{TX}}$ is the distance from the origin $O$ to the $\mathrm{P}_{\mathrm{TX}}$, i.e., the length of line segment $\overline{O \mathrm{P}_{\mathrm{TX}}}$. The DT is located at $\mathrm{P}_{\mathrm{DT}}$, with coordinates $(0,l_{\mathrm{DT}}\sin \gamma ,l_{\mathrm{DT}}\cos \gamma )$, where $l_{\text{DT}}$ is the distance from the origin $O$ to the $\mathrm{P}_{\mathrm{DT}}$, i.e., the length of line segment $\overline{O \mathrm{P}_{\mathrm{DT}}}$, $\gamma \in (-\pi/2,\pi/2)$ is the angle between the vector $\overrightarrow{O\mathrm{P}_{\mathrm{DT}}}$ and the positive half of the $z$-axis.\footnote{For simplicity, we only consider the scenario where the TX is in the boresight of the RIS. In this case, the position of DT can be described via the distance $l_{\mathrm{DT}}$ and the angle $\gamma$ by exploiting the symmetry of the RIS. Furthermore, in scenarios where the TX is not in the boresight of the RIS, our proposed method can also be readily adjusted to accommodate, which is illustrated in Section \ref{sec:RIS Wideband Beamforming Design}.} 

The circular RIS comprises $N_{\text{RIS}}$ reflecting elements with a radius of $R$. Furthermore, let $\Delta=\lambda_{\mathrm{c}}/2$ be the spacing between adjacent RIS elements along the $x$ and $y$ axis, where $\lambda_{\text{c}}=\frac{c}{f_{\text{c}}}$ is the wavelength at the center frequency $f_{\mathrm{c}}$ and $c$ is the speed of light. The set of 3D coordinates associated with RIS elements is defined as $\mathcal{S}_{\text{D}}$, which has $N_{\text{RIS}}$ coordinates that satisfy $x^2+y^2\leq R^2$ and $z=0$. Each RIS element is seen as a reconfigurable phase shifter, while the RIS is considered as a network with $N_{\text{RIS}}$ reconfigurable phase shifters.

\subsection{RIS Wideband Near-field Channel Model}
\label{subsec_system_model_B}
Next, we explain the RIS wideband channel in the near-field line of sight (LoS) scenario.\footnote{The system is considered near-field if the transceiver distance is smaller than the Fraunhofer distance when the EM wave cannot be simplified as the planar wave \cite{selvan2017fraunhofer_rayleigh_distance}. Since the considered RIS is equipped with massive reflecting elements, it is appropriate to make the near-field assumption.} Let $h(x,y,f)$ denote the cascaded channel between the TX and the DT via the RIS element $(x,y,0) \in \mathcal{S}_{\text{D}}$ at a frequency of $f$, which can be expressed as \cite{mmwave_channel_model}
\begin{align}
\label{cascaded_channel}
h(x,y,f)&=h_{\mathrm{TX}}\left( x,y,f \right) *h_{\mathrm{DT}}\left( x,y,f \right) 
\\
&=\frac{ce^{-j2\pi fl_{\mathrm{TX}}\left( x,y \right) /c}}{2\pi fl_{\mathrm{TX}}\left( x,y \right)}*\frac{ce^{-j2\pi fl_{\mathrm{DT}}\left( x,y \right) /c}}{2\pi fl_{\mathrm{DT}}\left( x,y \right)}
\\
&=\frac{c^2e^{\left\{ -j2\pi f\left[ l_{\mathrm{TX}}\left( x,y \right) +l_{\mathrm{DT}}\left( x,y \right) \right] \right\} /c}}{\left( 2\pi f \right) ^2l_{\mathrm{TX}}\left( x,y \right) l_{\mathrm{DT}}\left( x,y \right)},
\end{align}
where $j=\sqrt{-1}$, and $h_{\mathrm{TX}}\left( x,y,f \right) $/$h_{\mathrm{DT}}\left( x,y,f \right) $ is the channel between the RIS element located in $(x,y,0)$ and the TX/DT. In addition, $l_{\mathrm{TX}}\left( x,y \right)$ is the distance between the TX and the RIS element $(x,y,0)$, which is given by $l_{\mathrm{TX}}\left( x,y \right) =|\mathrm{P}_{x,y}-\mathrm{P}_{\mathrm{TX}}|=\sqrt{x^2+y^2+{l_{\mathrm{TX}}}^2}$, while $l_{\mathrm{DT}}\left( x,y \right)$ is the distance between the DT and the RIS element $(x,y,0)$, which is given by $ l_{\mathrm{DT}}\left( x,y \right) =|\mathrm{P}_{\mathrm{DT}}-\mathrm{P}_{x,y}|=\sqrt{x^2+\left( y-l_{\mathrm{DT}}\sin \gamma \right) ^2+\left( l_{\mathrm{DT}}\cos \gamma \right) ^2}$. Furthermore, we let $l_{\mathrm{TX}}\left( 0,0 \right) =l_{\mathrm{TX}}$ and $l_{\mathrm{DT}}\left( 0,0 \right)=l_{\mathrm{DT}}$ for simplicity.

With the phased array architecture, the RIS can control the RF signals transmitted from the TX by adjusting the phase shifts of its elements. Let $g\left( f \right)$ denote the frequency domain single-input single-output (SISO) channel between the TX and the DT, which is given by
\begin{equation}
\label{g_f}
    g\left( f \right) =\sum_{\left( x,y,0 \right) \in \mathcal{S} _\mathrm{D}}{e^{j\phi \left( x,y \right)} h\left( x,y,f \right)},
\end{equation}
where $\phi \left( x,y \right)$ is the phase shift applied to the RF signals at the RIS element $(x,y,0)$. And, the channel gain affected by the corresponding phase shifts is defined as $|g\left( f \right) |^2$.  

\subsection{Problem Formulation and Analysis}
\label{subsec_system_model_C}

\begin{figure}[t]
    \centering
    \includegraphics[width=0.7\linewidth]{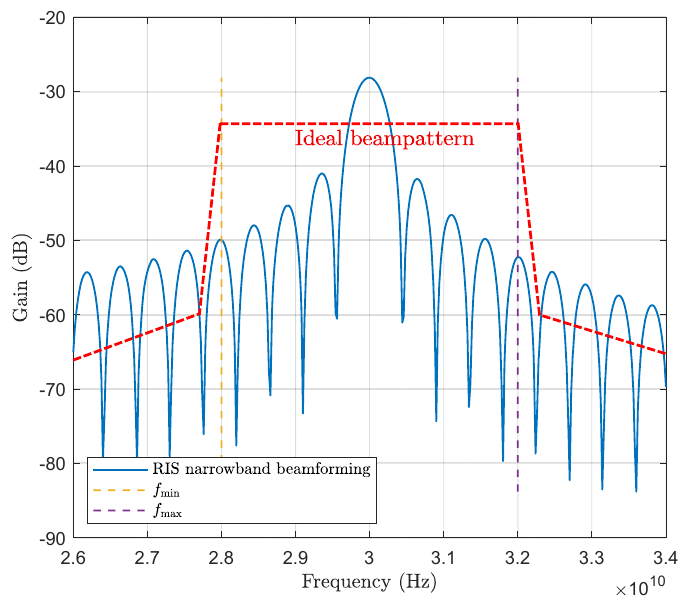}
    \caption{The ideal RIS wideband beampattern and the standard narrowband beampattern ($f_{\mathrm{c}}=30$ GHz, $B=4$ GHz, $R=1$ m, $l_{\mathrm{TX}}=0.5$ m, $l_{\mathrm{DT}}=10$ m).}
    \label{illustration_problem_formulation}
\end{figure}

Generally, the goal of the beamforming is to optimize the phase shift $\phi \left( x,y \right)$ to maximize the received power. For instance, in a narrowband system operating at the carrier frequency $f_{\mathrm{c}}$, standard narrowband beamforming adjusts the phase shift $\phi_{\mathrm{std}}(x,y)$ to maximize the channel gain $|g\left( f_{\mathrm{c}} \right) |^2$. The standard narrowband phase shift $\phi_{\mathrm{std}}(x,y)$ can be expressed as 
\begin{equation}
\label{standard_narrowband_phi}
    \phi _{\mathrm{std}}(x,y)=\frac{2\pi f_{\mathrm{c}}l\left( x,y \right)}{c},
\end{equation}
where $l\left( x,y \right)=l_{\mathrm{TX}}\left( x,y \right)+ l_{\mathrm{DT}}\left( x,y \right) $ is the sum of distances traveled by the signal as it propagates from the TX to the DT via the reflecting element $(x,y,0)$ of the RIS. Unfortunately, due to the frequency-independent constraint on $\phi \left( x,y \right)$, the RIS standard narrowband beamforming will cause dramatic gain variation over the desired wide bandwidth. As illustrated in Fig.\ref{illustration_problem_formulation}, the frequency-domain beampattern corresponding to the RIS standard narrowband beamforming is exemplified by the blue curve, which shows that the maximum gain and the minimum gain differ by more than 40 dB over a 4 GHz bandwidth.

However, an ideal wideband beamformer maximizes $|g\left( f \right)|^2$ at all frequencies within the desired bandwidth. The corresponding ideal beampattern is shown in Fig.\ref{illustration_problem_formulation} using the red dash lines, where the bandwidth $B=4$ GHz, the minimum desired frequency $f_{\min}=28$ GHz, and the maximum desired frequency $f_{\max}=32$ GHz. Since the channel in (\ref{cascaded_channel}) varies with the frequency, such an ideal beamformer must be frequency-dependent, which can not be realized due to hardware constraints in practice.

Taking into account practical applications, a typical wideband beamformer $\phi _{\mathrm{opt}}$ should have a large and approximately flat beampattern over the desired wide bandwidth \cite{myers2021infocus}, which can be mathematically expressed as 
\begin{equation}
\label{ideal_beamformer}
    \phi _{\mathrm{opt}}=\mathop {\mathrm{arg}\max} \limits_{\phi}\left\{ \min_f |g\left( f \right) |^2 \right\}, 
\end{equation}
where $g(f)$ depends on the phase shift $\phi$ through (\ref{g_f}). The optimization problem in (\ref{ideal_beamformer}) is non-convex. By using the semi-definite relaxation (SDR) method \cite{liu2018space_SDR}, the problem can be transformed into a convex optimization problem. However, the complexity of solving the problem through SDR is $\mathcal{O}(N^{4.5}_{\mathrm{RIS}})$ which is intolerable due to the massive number of RIS elements \cite{luo2010semidefinite}. \footnote{Typically, the number of RIS elements in mmWave bands is 256, 1024, 4096, or even more which causes unacceptable complexity for the convex optimization or the heuristic method.} 

Therefore, in the paper, we focus on the wideband beamforming of the phase-shifter-based RIS. Specifically, the objective is to design the phase shift $\phi$ such that $|g\left( f \right) |^2$ is large and approximately flat over the desired wide bandwidth. By combining the use of SFFT and SPM, instead of convex optimization, we provide a new way to address the RIS wideband beamforming problem with much lower computational complexity.

\section{SFFT-based RIS Wideband Beamforming Design}
\label{sec:RIS Wideband Beamforming Design}
In this section, we demonstrate the SFFT-based RIS wideband beamforming design. To simplify the analysis, in Subsection \ref{WBF_design_A} and \ref{WBF_design_B}, we approximate the discrete RIS into the continuous RIS and transform the RIS wideband beamforming problem into designing the additional phase shift $\phi_{\mathrm{des}}(x,y)$ to complement the existing standard narrowband phase shift $\phi_{\mathrm{std}}(x,y)$. Then, in Subsection \ref{WBF_design_C}, we design the additional phase shift $\phi_{\mathrm{des}}(x,y)$ via the SFFT and the SPM. For ease of understanding, we first solve the problem for the special scenario where the DT is along the boresight of the RIS. Then, we extend the special scenario to the general scenario where the location of DT is arbitrary.

\subsection{From the Discrete to the Continuous}
\label{WBF_design_A}
We approximate the discrete RIS as the continuous RIS by transforming the summation in (\ref{g_f}) into an integral because the integral is easier to deal with than the summation. The approximation considers a RIS with the continuous aperture of radius $R$ which contains an uncountable number of infinitely small reflecting elements. The set of the corresponding coordinates of the RIS elements is defined as $\mathcal{S}=\{\left( x,y,z \right)|x^2+y^2 \le R^2,z=0 \}$.\footnote{The concept of the continuous array, or rather, the holographic array, can be referred to in \cite{holographic_mimo_imaginary_surface}.} The idea behind the approximation is that we first design the phase shift $\phi(x,y)$ that achieves wideband beamforming in the continuous RIS case. Then, the phase shift $\phi(x,y)$ can be sampled from $\mathcal{S}$ to $\mathcal{S_{\mathrm{D}}}$ for a discrete RIS with $N_{\mathrm{RIS}}$ reflecting elements. 

For a continuous RIS, the phase shift $\phi(x,y)$ can be applied to control the reflection of the RF signals by manipulating infinitely small reflecting elements of the RIS. Analogous to $g(f)$ in (\ref{g_f}), the corresponding RIS SISO channel is
\begin{align}
g_{\mathrm{a}}(f)&=\int_{\mathcal{S}}{h}\left( x,y,f \right) e^{j\phi \left( x,y \right)}\mathrm{d}x\mathrm{d}y
\label{g_a_f}
\\
&=\frac{c^2}{4\pi ^{2}f^2}\int_{\mathcal{S}}{\frac{e^{j\left\{ \phi \left( x,y \right) -\frac{2\pi f\left[ l_{\mathrm{TX}}\left( x,y \right) +l_{\mathrm{DT}}\left( x,y \right) \right]}{c} \right\}}}{l_{\mathrm{TX}}\left( x,y \right) l_{\mathrm{DT}}\left( x,y \right)}\mathrm{d}x\mathrm{d}y},
\label{g_a_f_l_TX_l_DT}
\end{align}
where $g_{\mathrm{a}}(f)$ is interpreted as a continuous approximation, i.e., the Riemann sum approximation, of $g(f)$ in (\ref{g_f}). The channel $g_{\mathrm{a}}(f)$ in an integral format enables the closed-form solution for the wideband beamforming problem. 

\subsection{From the Narrowband to the Wideband}
\label{WBF_design_B}
Instead of directly designing the wideband phase shift $\phi(x,y)$, we transform the problem into designing the additional phase shift $\phi_{\mathrm{des}}(x,y)$ to complement the existing standard narrowband phase shift $\phi_{\mathrm{std}}(x,y)$. Adding $\phi_{\mathrm{des}}(x,y)$ on $\phi_{\mathrm{std}}(x,y)$ given in (\ref{standard_narrowband_phi}) yields the wideband phase shift 
\begin{equation}
\label{phi_added}
    \phi(x,y)=\phi_{\mathrm{std}}(x,y)+\phi_{\mathrm{des}}(x,y).
\end{equation}

By substituting (\ref{phi_added}) into (\ref{g_a_f_l_TX_l_DT}), the channel $g_{\mathrm{a}}(f)$ can be expressed as
\begin{align}
 g_{\mathrm{a}}(f)&=\eta \left( f \right) \int_{\mathcal{S}}{\frac{e^{j\left\{ \phi _{\mathrm{std}}(x,y)+\phi _{\mathrm{des}}(x,y)-\frac{2\pi fl\left( x,y \right)}{c} \right\}}}{l_{\mathrm{TX}}\left( x,y \right) l_{\mathrm{DT}}\left( x,y \right)}\mathrm{d}x\mathrm{d}y}
\\
&=\eta \left( f \right) \int_{\mathcal{S}}{\frac{e^{j\phi _{\mathrm{des}}(x,y)}e^{-j\frac{2\pi \left( f-f_{\mathrm{c}} \right) l\left( x,y \right)}{c}}}{l_{\mathrm{TX}}\left( x,y \right) l_{\mathrm{DT}}\left( x,y \right)}\mathrm{d}x\mathrm{d}y}
\\
&=\eta \left( f \right) \int_{\mathcal{S}}{\frac{e^{j\phi _{\mathrm{des}}(x,y)}e^{-j\omega l(x,y)}}{l_{\mathrm{TX}}\left( x,y \right) l_{\mathrm{DT}}\left( x,y \right)}\mathrm{d}x\mathrm{d}y},
\label{g_a_f_with_eta}
\end{align}
where $w=\frac{2\pi \left( f-f_{\mathrm{c}} \right)}{c}$ is the spatial frequency with units of radians/meter (rad/m), $\eta \left( f \right) =\frac{c^2}{4\pi ^{2}f^2}$ is the amplitude coefficient which is dependent on frequency $f$, and $l(x,y)=l_{\mathrm{TX}}\left( x,y \right) +l_{\mathrm{DT}}\left( x,y \right)$ is the sum of distances traveled by the signal as it propagates from the TX to the DT via the RIS reflecting element $(x,y,0)$.

The question here is to determine a 2D phase function $\phi _{\mathrm{des}}(x,y)$ that enables a large and flat gain $|g_{\mathrm{a}}(f)|^2$ over $[f_{\mathrm{c}}-B/2,f_{\mathrm{c}}+B/2]$. For a tractable design of phase shifts, we ignore the coefficient $\eta \left( f \right)$ in (\ref{g_a_f_with_eta}) and define 
\begin{align}
\label{tilde_g_a_f_without_eta}
    \tilde{g}_{\mathrm{a}}(f)=\int_{\mathcal{S}}{\frac{e^{j\phi _{\mathrm{des}}(x,y)}e^{-jwl\left( x,y \right)}}{l_{\mathrm{TX}}\left( x,y \right) l_{\mathrm{DT}}\left( x,y \right)}\mathrm{d}x\mathrm{d}y}.
\end{align}
The RIS wideband beamforming aims to design $\phi _{\mathrm{des}}(x,y)$ such that $|\tilde{g}_{\mathrm{a}}(f)|^2$ is large and approximately flat over the desired bandwidth. We ignore the $1/f^2$ term, i.e., $\eta(f)$, in $g_{\mathrm{a}}(f)$ because it leads to a much smaller variation in $|g_{\mathrm{a}}(f)|^2$ compared to the integral term in (\ref{g_a_f_with_eta}). For example, the variation in  $|g_{\mathrm{a}}(f)|^2$ due to the $1/f^2$ term is about 1.2 dB for a 4 GHz system at 30 GHz. However, the variation in $|g_{\mathrm{a}}(f)|^2$ due to $\tilde{g}_{\mathrm{a}}(f)$ can be more than 40 dB with the RIS standard narrowband beamforming design as shown in Fig.\ref{illustration_problem_formulation}.

\subsection{From the Space Domain to the Frequency Domain}
\label{WBF_design_C}
In this subsection, we solve the problem of the RIS wideband beamforming by exploiting SFFT and SPM. Essentially, the formula (\ref{tilde_g_a_f_without_eta}) reveals how the spatial structure affects the frequency-domain signal via phase shifts. Inspired by this, we further analyze the space-frequency relationship to solve the RIS wideband beamforming problem. For ease of understanding, we divide this subsection into two parts. In the first part, we discuss the special scenario where the DT is along the boresight. In the second part, we extend the discussion into the general scenario where the location of the DT is arbitrary.

\subsubsection{\textbf{RIS Wideband Beamforming When the Location of the DT is Along the Boresight}} 

\begin{figure}
    \centering
    \includegraphics[width=0.6\linewidth]{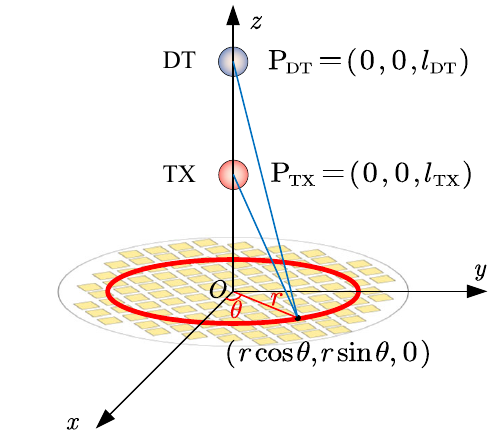}
    \caption{The scenario when the DT is along the boresight.}
    \label{fig:boresight_case}
\end{figure}

We begin from the special case where the DT is along the boresight of the RIS ($ \gamma =0 ^\circ$). In this case, the TX and the DT are both in the positive half of the $z$-axis of $\mathcal{C}_0$. Taking advantage of the system's symmetry, we introduce the polar coordinate representation. Let $x=r\mathrm{cos}\theta$, and $y=r\mathrm{sin}\theta$, where $r= \sqrt{x^2 + y^2} \in[0,R]$ is the radial distance and $\theta=\arctan(\frac{y}{x})\in[0,2\pi)$ is the polar angle as shown in Fig.\ref{fig:boresight_case}. 
Substituting $(x,y)$ by polar coordinates $(r,\theta)$, we have $\phi_{\mathrm{des}}(x,y)=\phi_{\mathrm{des}}(r \mathrm{cos}\theta,r \mathrm{sin} \theta)$, $l_{\mathrm{TX}}\left( x,y \right)=l_{\mathrm{TX}}\left( r \right)=\sqrt{{l_\mathrm{TX}}^2+r^2} $, $l_{\mathrm{DT}}\left( x,y \right)=l_{\mathrm{DT}}\left( r \right)=\sqrt{{l_\mathrm{DT}}^2+r^2}$, and $l\left( x,y \right)=l(r)=\sqrt{{l_\mathrm{TX}}^2+r^2}+\sqrt{{l_\mathrm{DT}}^2+r^2}$, where $l_{\mathrm{TX}}(r)$, $l_{\mathrm{DT}}(r)$, and $l(r)$ are independent of $\theta$ due to the radial symmetry in the boresight scenario. 
{\color{black}Moreover, the RIS elements with the same $l(r)$ should have identical phase shifts. Therefore, due to the radial symmetry, the additional phase shift $\phi _{\mathrm{des}}\left( r\cos \theta ,r\sin \theta \right) $ varies only with $r$ and is independent of the $\theta$, i.e., $\phi _{\mathrm{des}}\left( x,y \right) =\phi _{\mathrm{des}}\left( r \right) $, $\forall \sqrt{x^2+y^2}=r$. }

As such, $\tilde{g}_{\mathrm{a}}(f)$ in (\ref{tilde_g_a_f_without_eta}) can be transformed into 
\begin{align}
\label{tilde_g_a_f_polar}
\tilde{g}_{\mathrm{a}}(f)=\int_{\theta =0}^{2\pi}{\int_{r=0}^R{\frac{e^{j\phi _{\mathrm{des}}\left( r \right)}e^{-j\omega l\left( r \right)}}{\sqrt{{l_{\mathrm{TX}}}^2+r^2}\sqrt{{l_{\mathrm{DT}}}^2+r^2}}r\mathrm{d}r}\mathrm{d}\theta}.
\end{align}
Recall $l(r)=\sqrt{{l_\mathrm{TX}}^2+r^2}+\sqrt{{l_\mathrm{DT}}^2+r^2}$, and we have
\begin{align}
\label{dl}
    \mathrm{d}l=\frac{r\mathrm{d}r\left( \sqrt{{l_{\mathrm{TX}}}^2+r^2}+\sqrt{{l_{\mathrm{DT}}}^2+r^2} \right)}{\sqrt{{l_{\mathrm{TX}}}^2+r^2}\sqrt{{l_{\mathrm{DT}}}^2+r^2}},
\end{align}
which can be further rewritten as
\begin{align}
\label{rdr}
    \frac{r\mathrm{d}r}{\sqrt{{l_{\mathrm{TX}}}^2+r^2}\sqrt{{l_{\mathrm{DT}}}^2+r^2}}=\frac{\mathrm{d}l}{l}.
\end{align}
By substituting (\ref{rdr}) into (\ref{tilde_g_a_f_polar}), the equation (\ref{tilde_g_a_f_polar}) can be simplified as 
\begin{align}
\label{g_a_f_substitution_l}
\tilde{g}_{\mathrm{a}}(f)&=\int_{\theta =0}^{2\pi}{\int_{l=l_{\min}}^{l_{\max}}{\frac{1}{l}e^{j\phi _{\mathrm{des}}\left( l \right)}e^{-j\omega l}\mathrm{d}l}\mathrm{d}\theta}
\\
&=\int_{\theta =0}^{2\pi}{\mathrm{d}\theta \int_{l=l_{\min}}^{l_{\max}}{\frac{1}{l}e^{j\phi _{\mathrm{des}}\left( l \right)}e^{-j\omega l}\mathrm{d}l}}
\\
&=2\pi \int_{l=l_{\min}}^{l_{\max}}{\frac{1}{l}e^{j\phi _{\mathrm{des}}\left( l \right)}e^{-j\omega l}\mathrm{d}l},\label{1D_g_a_f}
\end{align}
where $l_{\min}=l_{\mathrm{TX}}+l_{\mathrm{DT}}$, $l_{\max}=\sqrt{{l_{\mathrm{TX}}}^2+R^2}+\sqrt{{l_{\mathrm{DT}}}^2+R^2}$, and $\phi _{\mathrm{des}}\left( l \right)$ is a one-dimentional (1D) function. By doing the above transformation, the problem is simplified into designing a 1D function $\phi_{\mathrm{des}}(l)$. Let $a(l)=\frac{1}{l}$ and ignore the constant scaling $2\pi$, the equation (\ref{1D_g_a_f}) can be further rewritten as
\begin{align}
\label{1D_g_a_f_a_l}
\tilde{g}_{\mathrm{a}}(f)=\int_{l=l_{\min}}^{l_{\max}}{a\left( l \right) e^{j\phi _{\mathrm{des}}\left( l \right)}e^{-j\omega l}\mathrm{d}l}.
\end{align}
Next, we replace the frequency $f$ with the spatial frequency $\omega$, where $\omega=\frac{2\pi \left( f-f_{\mathrm{c}} \right)}{c}\in[-\pi B/c,\pi B/c]$, and define $\mathbb{I} _{l_{min}}^{l_{\max}}\left( l \right) $ as the indicator function which is equal to 1 for $l \in [l_{\min},l_{\max}]$ and is equal to 0 otherwise. As such, the equation (\ref{1D_g_a_f_a_l}) can be expressed as 
\begin{align}
\hat{g}_a\left( \omega \right) =\int_{-\infty}^{\infty}{a\left( l \right) e^{j\phi_{\mathrm{des}} \left( l \right)}\mathbb{I} _{l_{\min}}^{l_{\max}}\left( l \right) e^{-j\omega l}\mathrm{d}l}.
\label{g_a_f_hat_a_l_FT_form}
\end{align}

To get more insight into the above mathematical expression (\ref{g_a_f_hat_a_l_FT_form}), we can heuristically assume that the term $a(l)$ in the equation (\ref{g_a_f_hat_a_l_FT_form}) is a constant. For example, if we let  $a(l)=1$, then the equation (\ref{g_a_f_hat_a_l_FT_form}) can be simplified as 
\begin{align}
\label{g_a_f_hat_FT_form}
\hat{g}_{\mathrm{a}}(\omega) =\int_{-\infty}^{\infty}{ e^{j\phi_{\mathrm{des}} \left( l \right)}\mathbb{I} _{l_{\min}}^{l_{\max}}\left( l \right) e^{-j\omega l}\mathrm{d}l}.
\end{align}
From equation (\ref{g_a_f_hat_FT_form}), we can observe that $\hat{g}_{\mathrm{a}}(\omega)$ is a Fourier transform of $e^{j\phi_{\mathrm{des}} \left( l \right)}\mathbb{I} _{l_{\max}}^{l_{\max}}\left( l \right) $. The equation (\ref{g_a_f_hat_FT_form}) mathematically reconstructs a space-domain function into a frequency-domain function. Inspired by this, we can use the properties of the Fourier transform to deal with the RIS wideband beamforming problem. 
Specifically. in the time-frequency Fourier transform, a linear frequency modulation continuous wave (LFM-CW) in the time domain is proven to have a relatively large and flat gain over particular frequency bands in the frequency domain \cite{levanon2004radar}. Therefore, we can design the additional phase shift $\phi_{\mathrm{des}}(l)$ in a similar mathematical expression as LFM-CW to enable a large and flat gain $|\hat{g}_{\mathrm{a}}(\omega)|^2$ over $[-\pi B/c,\pi B/c]$ based on equation (\ref{g_a_f_hat_FT_form}).

After we get the insight that the equation (\ref{g_a_f_hat_a_l_FT_form}) is an SFFT, we can easily find that $a(l)$ in the equation (\ref{g_a_f_hat_a_l_FT_form}) is actually an amplitude modulation function. To design $\phi_{\mathrm{des}}(l)$, we introduce the SPM \cite{myers2021infocus,levanon2004radar}, which characterizes the impact of this amplitude modulation on $\hat{g}_{\mathrm{a}}(\omega)$ under the assumption that $a(l)$ varies slowly when compared to $\phi_{\mathrm{des}} \left( l \right)$. Let $\omega_{l}$ denote the instantaneous frequency of $e^{j\phi_{\mathrm{des}} \left( l \right)}\mathbb{I} _{l_{\min}}^{l_{\max}}\left( l \right) e^{-j\omega l}$ at $l$, i.e., $\omega_{l}=\phi_{\mathrm{des}} '\left( l \right) $, where $\phi_{\mathrm{des}} '\left( l \right)$ is the first order derivative of $\phi_{\mathrm{des}}\left( l \right)$. The SPM approximates $|\hat{g}_{\mathrm{a}}(\omega_l)|^2$ as 
\begin{align}
   |\hat{g}_{\mathrm{a}}(\omega _l)|^2\approx \frac{2\pi a^2(l)}{|\phi_{\mathrm{des}} ''\left( l \right) |},
    \label{stationary_phase_method_SPM}
\end{align}
where $\phi_{\mathrm{des}} ''\left( l \right)$ is the second order derivative of $\phi_{\mathrm{des}}\left( l \right)$. Since $\omega_l$ is a continuous function that increases from $\phi_{\mathrm{des}} '\left( l_{\mathrm{min}} \right)=-\pi B/c$ to $\phi_{\mathrm{des}} '\left( l_{\mathrm{max}} \right)=\pi B/c$, we can infer that $\phi_{\mathrm{des}} ''\left( l \right)>0,~\forall l \in [l_{\mathrm{min}},l_{\mathrm{max}}]$. However, the increase can be non-linear, which depends on $a(l)$. Considering the condition that $\phi_{\mathrm{des}} ''\left( l \right)>0$, the relationship in (\ref{stationary_phase_method_SPM}) can be further expressed as
\begin{align}
   |\hat{g}_{\mathrm{a}}(\omega_l )|^2 \approx \frac{2\pi a^2(l)}{ \phi_{\mathrm{des}} ''\left( l \right)}.
    \label{phi_dd}
\end{align}
It can be inferred from (\ref{phi_dd}) that the phase shift $ \phi_{\mathrm{des}}\left( l \right)$ that achieves a flat $|\hat{g}_{\mathrm{a}}(\omega )|^2$ over $\omega \in [-\pi B/c,\pi B/c]$ satisfy
\begin{align}
    \phi_{\mathrm{des}} ''\left( l \right)=\kappa a^2(l),
    \label{phi_dd_kappa}
\end{align}
where $\kappa$ is a positive constant. 

Then, by using $\phi_{\mathrm{des}} '(l)=\int_{l_{\min}}^l{\phi_{\mathrm{des}} ''\left( l \right)}\mathrm{d}l$, where $\phi_{\mathrm{des}} '(l_{\min})=-\pi B/c$ and $\phi_{\mathrm{des}} '(l_{\max})=\pi B/c$, $\phi_{\mathrm{des}} '(l)$ can be derived as
\begin{align}
    \phi_{\mathrm{des}} '(l)=\frac{2\pi B\int_{l_{\min}}^l{a^2\left( l \right) \mathrm{d}l}}{c\int_{l_{\min}}^{l_{\max}}{a^2\left( l \right) \mathrm{d}l}}-\frac{\pi B}{c},
    \label{phi_d}
\end{align}
where the integral of $a^2(l)$ in (\ref{phi_d}) can be computed through numerical integration. And we can derive the phase shift $\phi_{\mathrm{des}} (l)$ from (\ref{phi_d}) in the boresight scenario via
\begin{align}
    \phi_{\mathrm{des}} \left( l \right) =\int_{l_{\min}}^l{\phi_{\mathrm{des}} '(l)\mathrm{d}l},
    \label{phi_solution}
\end{align}
which can be computed using numerical integration, too.

Furthermore, $\phi_{\mathrm{des}}(x,y)$ can be calculated from $\phi_{\mathrm{des}}(l)$ via geometric relationship. Specifically, for $\forall l=l_{\mathrm{TX}}\left( x,y \right) +l_{\mathrm{DT}}\left( x,y \right) ,\left( x,y,0 \right) \in \mathcal{S} _{\mathrm{D}}$, we have
\begin{align}
    \phi_{\mathrm{des}} \left( x,y \right) =\phi_{\mathrm{des}} \left( l \right).
    \label{equ:phi_x_y_phi_l}
\end{align}

\begin{figure*}[t] % Use figure* for full-width figure across two columns
\centering
\begin{subfigure}[t]{0.28\linewidth}
    \centering
    \includegraphics[width=\linewidth,height=\linewidth]{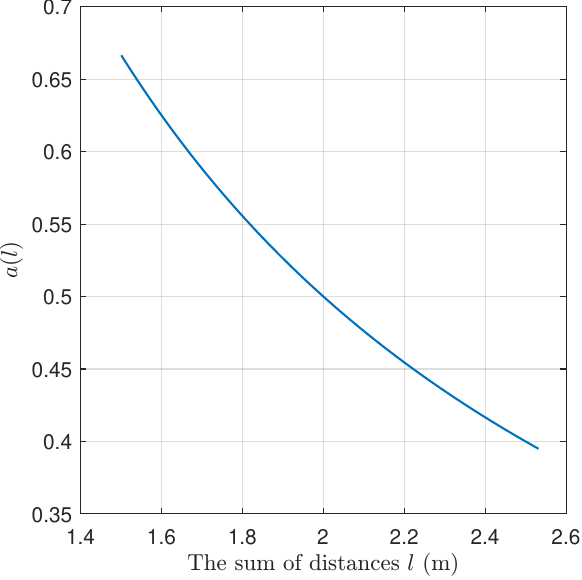}
    \caption{Modulation function $a(l)$.}
    \label{fig:amplitude-factor}
\end{subfigure}\hfil
\begin{subfigure}[t]{0.28\linewidth}
    \centering
    \includegraphics[width=\linewidth,height=\linewidth]{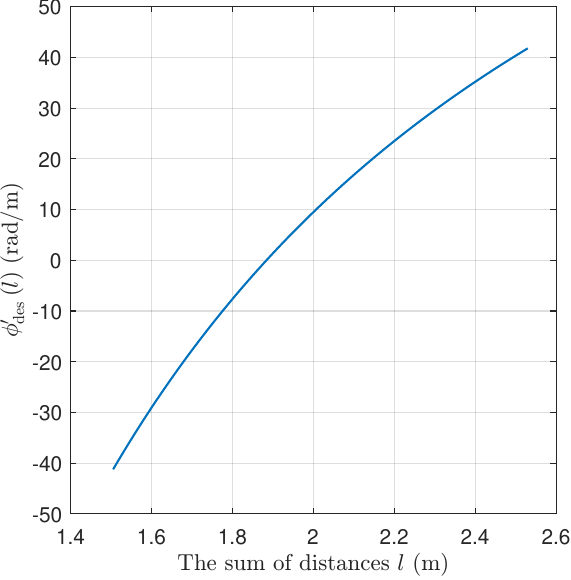}
    \caption{Instantaneous frequency $\phi_{\mathrm{des}} ' (l)$.}
    \label{fig:phi-prime-l}
\end{subfigure}\hfil
\begin{subfigure}[t]{0.28\linewidth}
    \centering
    \includegraphics[width=\linewidth,height=\linewidth]{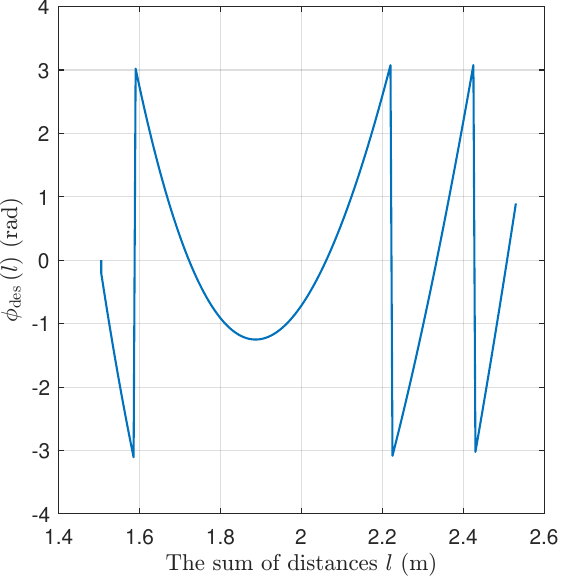}
    \caption{Additional phase shift $\phi_{\mathrm{des}} (l)$.}
    \label{fig:phi-l}
\end{subfigure}

\caption{Results of SPM in the boresight scenario.}
\label{fig:results of boresight SPM illstruation}
\end{figure*}

\begin{figure*}[t] % Use figure* for full-width figure across two columns
\centering
\begin{subfigure}[t]{0.33\linewidth}
    \centering
    \includegraphics[width=\linewidth]{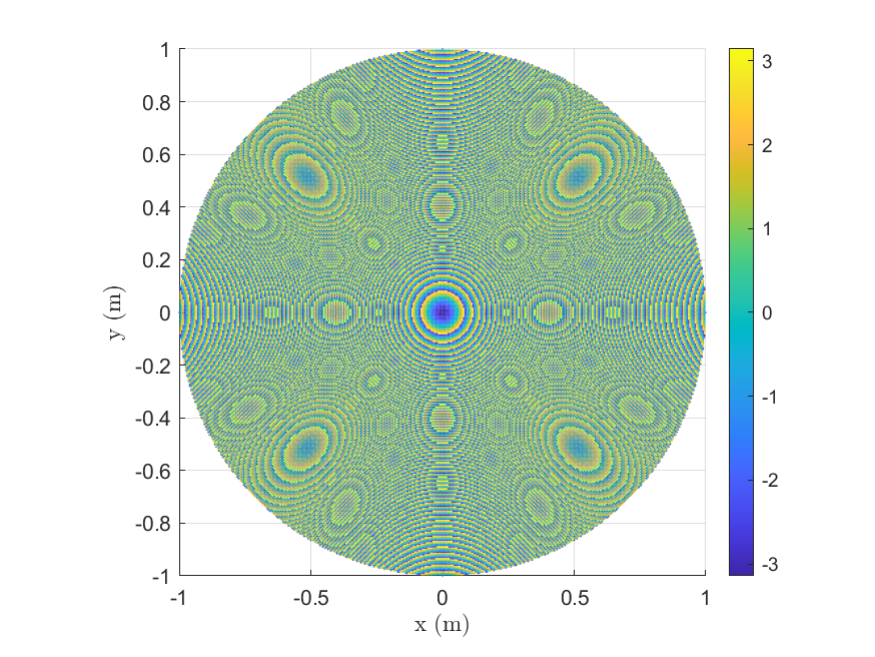}
    \caption{$\phi_{\mathrm{std}}(x,y)$.}
    \label{fig:phase_shift_boresight_a}
\end{subfigure}\hfil
\begin{subfigure}[t]{0.33\linewidth}
    \centering
    \includegraphics[width=\linewidth]{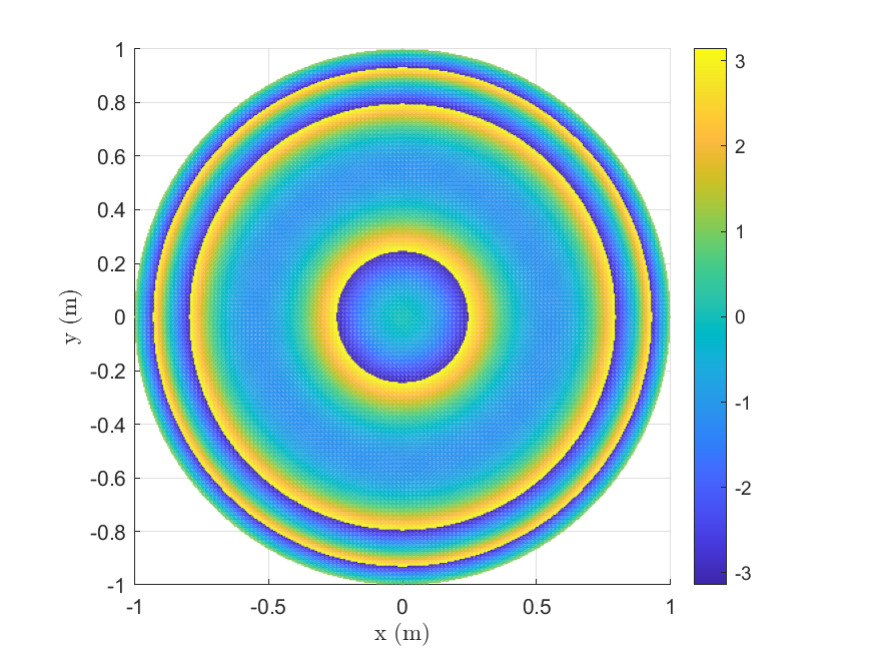}
    \caption{$\phi_{\mathrm{des}}(x,y)$.}
    \label{fig:phase_shift_boresight_b}
\end{subfigure}\hfil
\begin{subfigure}[t]{0.33\linewidth}
    \centering
    \includegraphics[width=\linewidth]{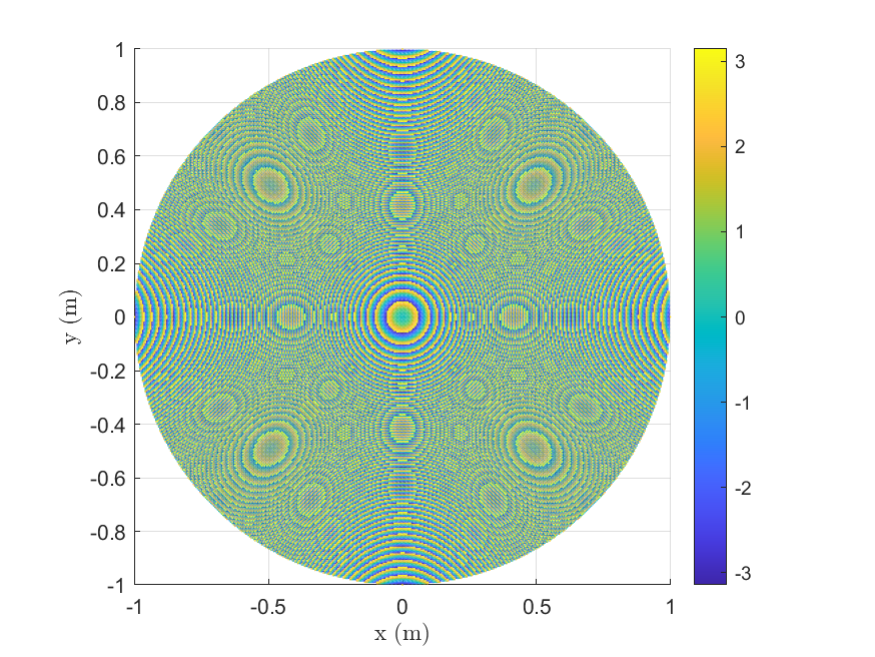}
    \caption{$\phi(x,y)=\phi_{\mathrm{std}}(x,y)+\phi_{\mathrm{des}}(x,y)$.}
    \label{fig:phase_shift_boresight_c}
\end{subfigure}

\caption{Boresight phase shifts.}
\label{fig:results of boresight phase shift}
\end{figure*}

Finally, by adding the solution $\phi_{\mathrm{des}}(x,y)$ in (\ref{equ:phi_x_y_phi_l}) and the standard narrowband phase shift $\phi_{\mathrm{std}}(x,y)$ in (\ref{standard_narrowband_phi}) via (\ref{phi_added}), we have the final solution $ \phi \left( x,y \right)$ for the RIS wideband beamforming in the boresight scenario. 

\begin{figure*}[t] % Use figure* for full-width figure across two columns
\centering
\begin{subfigure}[t]{0.28\linewidth}
    \centering
    \includegraphics[width=\linewidth]{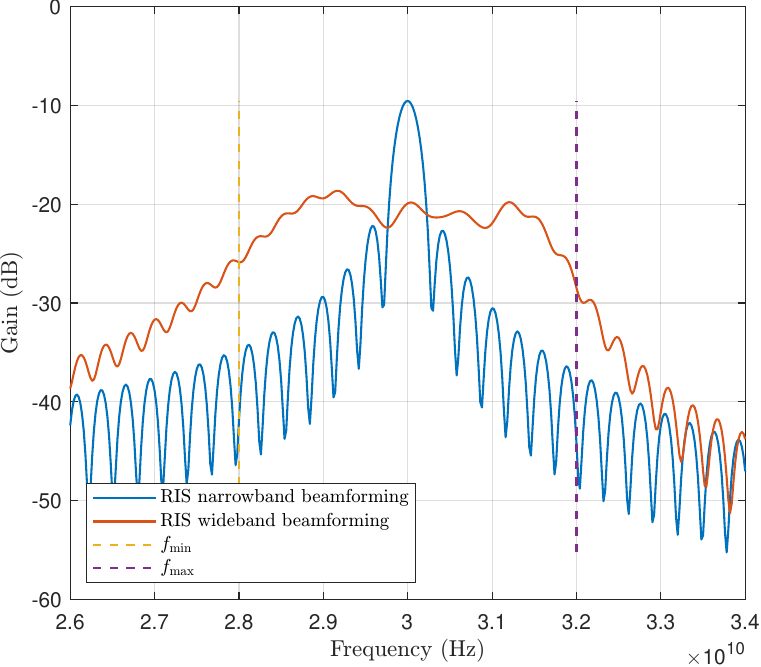}
    \caption{$l_{\mathrm{DT}}=1~$m.}
    \label{fig:beampattern_boresight_a}
\end{subfigure}\hfil
\begin{subfigure}[t]{0.28\linewidth}
    \centering
    \includegraphics[width=\linewidth]{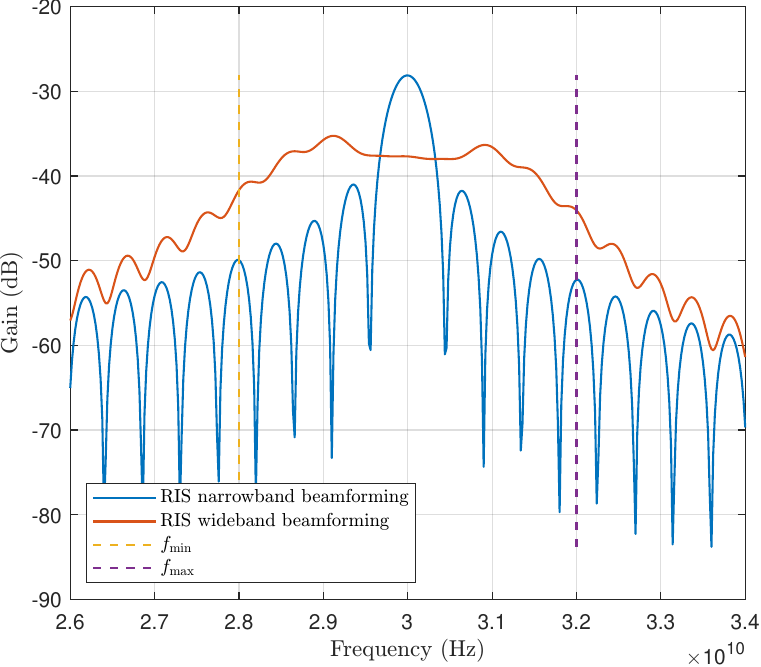}
    \caption{$l_{\mathrm{DT}}=10~$m.}
    \label{fig:beampattern_boresight_b}
\end{subfigure}\hfil
\begin{subfigure}[t]{0.28\linewidth}
    \centering
    \includegraphics[width=\linewidth]{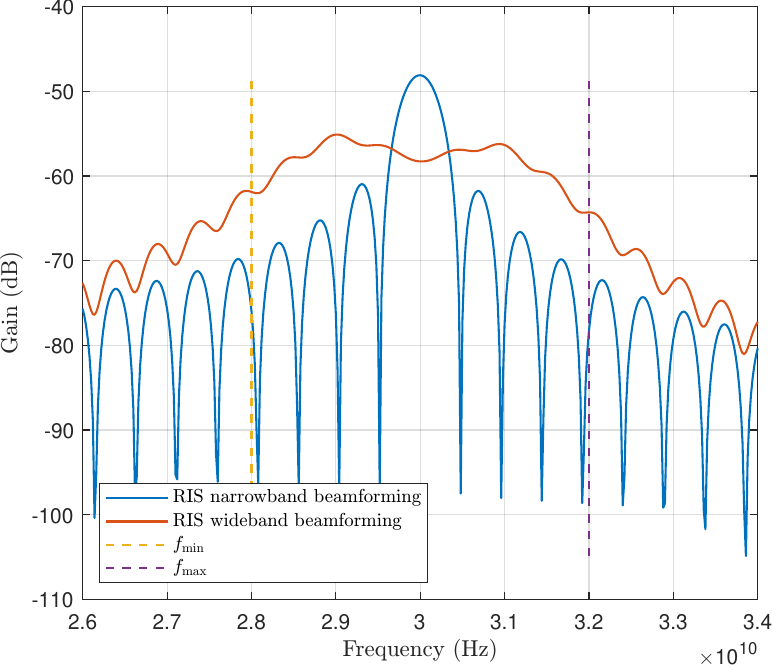}
    \caption{$l_{\mathrm{DT}}=100~$m.}
    \label{fig:beampattern_boresight_c}
\end{subfigure}

\caption{Boresight Beampatterns.}
\label{fig:results of boresight beampattern}
\end{figure*}

To show the effectiveness of the proposed method, we consider a system setup with parameters as follows: $f_{\mathrm{c}}=30$ GHz, $B=4$ GHz, $R=1$ m, $l_{\mathrm{TX}}=0.5$ m, $l_{\mathrm{DT}}=1$ m, and $\gamma_{\mathrm{c}}=0^\circ$.\footnote{Note that the simulation results shown in this paper are for the discrete RIS, while the analysis is for the continuous RIS.} The amplitude modulation function $a(l)$, the instantaneous frequency $\phi_{\mathrm{des}} '(l)$, and the designed additional phase shift $\phi_{\mathrm{des}}(l)$ are illustrated in Fig. \ref{fig:amplitude-factor}, Fig. \ref{fig:phi-prime-l} and Fig. \ref{fig:phi-l}, respectively. As shown in Fig.\ref{fig:results of boresight SPM illstruation}, the non-uniform nature of $a^2(l)=\frac{1}{l^2}$ over $[l_{\min},l_{\max}]$ results in the non-linear $\phi_{\mathrm{des}}'(l)$. The phase shift of standard narrowband beamforming $\phi_{\mathrm{std}}(x,y)$, the designed additional phase shift $\phi_{\mathrm{des}}(x,y)$, and the phase shift of the proposed wideband beamforming $\phi(x,y)$ are shown in Fig.\ref{fig:phase_shift_boresight_a}, Fig.\ref{fig:phase_shift_boresight_b}, and Fig.\ref{fig:phase_shift_boresight_c}, respectively. In Fig.\ref{fig:results of boresight beampattern}, we show the frequency-domain beampatterns of the proposed wideband beamforming and standard narrowband beamforming with different $l_{\mathrm{DT}}$. The proposed wideband beamforming method achieves relatively large and flat gain across the desired frequency range in all three $l_{\mathrm{DT}}$ configurations, indicating the significant potential of the proposed RIS wideband beamforming for facilitating wideband communication and sensing. A more detailed analysis of the performance of the proposed wideband beamforming method is provided in Section.\ref{sec:simulation results}.

\subsubsection{\textbf{RIS Wideband Beamforming When the Location of the DT is Arbitrary}}
We extend the derivation from the boresight case to the more general case where the location of the DT is arbitrary. We notice that the sum of distances $l$ is a critical variable. As shown in equation (\ref{standard_narrowband_phi}), (\ref{phi_added}) and (\ref{equ:phi_x_y_phi_l}), RIS elements with the same $l$ should have identical phase shift $\phi_{\mathrm{des}}(l)$, $\phi_{\mathrm{std}}(l)$, and $\phi(l)$. This observation enlightens us to further explore the geometric properties of the system to solve the RIS wideband beamforming problem. 

\begin{figure*}[t] % Use figure* for full-width figure across two columns
\centering
\begin{subfigure}[t]{0.28\linewidth}
    \centering
    \includegraphics[width=\linewidth,height=\linewidth]{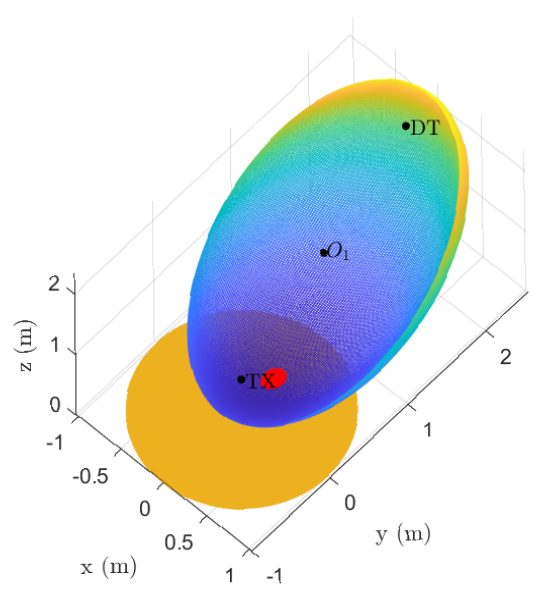}
    \caption{The intersection is a point.}
    \label{fig:intersection_a}
\end{subfigure}\hfil
\begin{subfigure}[t]{0.28\linewidth}
    \centering
    \includegraphics[width=\linewidth,height=\linewidth]{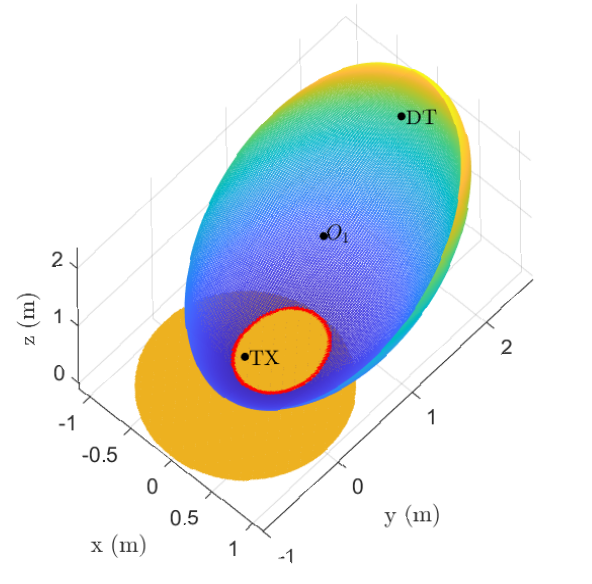}
    \caption{The intersection is an ellipse.}
    \label{fig:intersection_b}
\end{subfigure}
\begin{subfigure}[t]{0.28\linewidth}
    \centering
    \includegraphics[width=\linewidth,height=\linewidth]{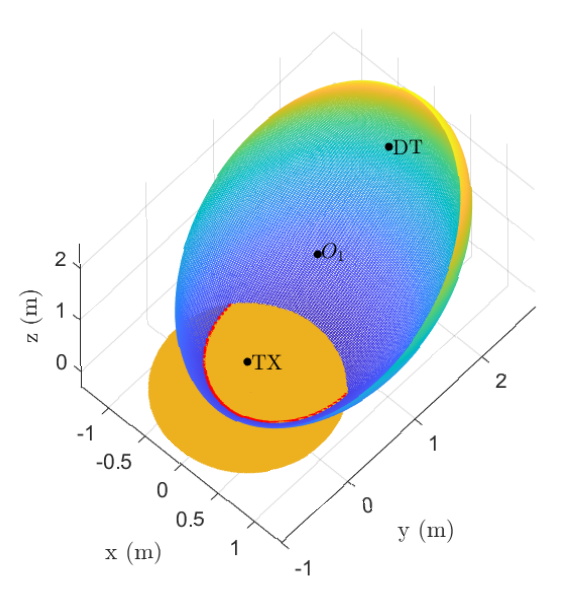}
    \caption{The intersection is an elliptical arc.}
    \label{fig:intersection_c}
\end{subfigure}\hfil
\caption{The Intersection $\mathcal{S}^{\mathrm{S}}_{\mathrm{RIS}}(l)$ generated by the spheroid and the RIS.}
\label{fig:intersection}
\end{figure*}

{\color{black}
Interestingly, the set of points with an equal sum of distances to two points is a spheroid in 3D space. In the proposed model, the geometric property of the equal sum of distances between points $\mathrm{P}_{\mathrm{TX}}$ and $\mathrm{P}_{\mathrm{DT}}$ can be precisely described by a prolate spheroid with foci $\mathrm{P}_{\mathrm{TX}}$ and $\mathrm{P}_{\mathrm{DT}}$. The prolate spheroid, which is a quadric surface, can be generated by rotating an ellipse about its major axes whose foci are $\mathrm{P}_{\mathrm{TX}}$ and $\mathrm{P}_{\mathrm{DT}}$.
}

We use $\mathcal{S}^{\mathrm{S}}_{z}(l)$ to denote the intersection of the spheroid and the $z=0$ plane and use $\mathcal{S}^{\mathrm{S}}_{\mathrm{RIS}}(l)$ to denote the intersection of the spheroid with given $l$ and the RIS. Obviously, the intersection $\mathcal{S}^{\mathrm{S}}_{z}(l)$ is an ellipse, a single point, or empty. Hence, the intersection $\mathcal{S}^{\mathrm{S}}_{\mathrm{RIS}}(l)$ is an ellipse, a segment of an elliptic arc, a single point, or empty, as shown in Fig. \ref{fig:intersection}. All RIS elements on the intersection $\mathcal{S}^{\mathrm{S}}_{\mathrm{RIS}}(l)$ should have the same phase shifts. As $l$ increases, the intersection $\mathcal{S}^{\mathrm{S}}_{\mathrm{RIS}}(l)$ transforms from a single point to an ellipse, gradually forming into an elliptical arc that will eventually vanish.

\begin{figure}[t] % Use figure* for full-width figure across two columns
\centering
\begin{subfigure}[t]{0.5\linewidth}
    \centering
    \includegraphics[width=\linewidth,height=\linewidth]{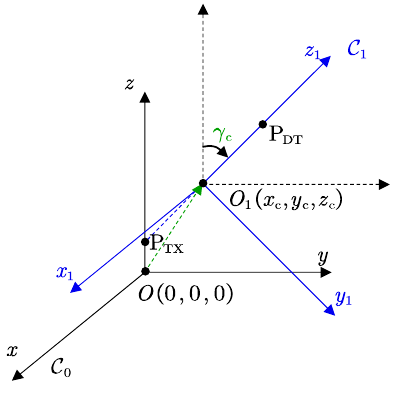}
    \caption{From $\mathcal{C}_0$ to $\mathcal{C}_1$.}
    \label{fig:coordinate_c0_to_c1}
\end{subfigure}\hfil
\begin{subfigure}[t]{0.5\linewidth}
    \centering
    \includegraphics[width=\linewidth,height=\linewidth]{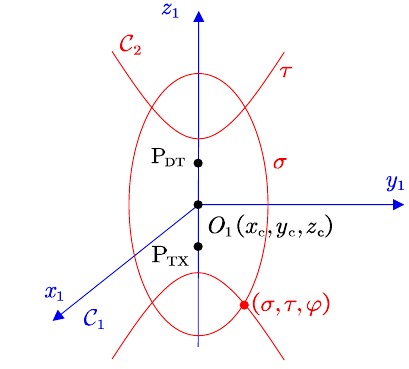}
    \caption{From $\mathcal{C}_1$ to $\mathcal{C}_2$.}
    \label{fig:coordinate_c1_to_c2}
\end{subfigure}\hfil
\caption{Coordinate system transformation.}
\label{fig:coordinate_system_trans}
\end{figure}

Next, we utilize the above geometric properties to solve the RIS wideband beamforming problem when the location of the DT is arbitrary.\footnote{Note that the boresight case is a special case where the intersection of the spheroid and $z=0$ plane is typically a circle rather than an ellipse which is much easier to deal with. We start with the special case and then extend to the general case because it is easier to understand.} Due to the complexity of the system's geometric properties, it is impossible to characterize the problem using a 2D polar coordinate system and reduce the dimensions of the integral variable as we did in the boresight case.
Nevertheless, we notice that the system can be modeled using the prolate spheroidal coordinate system. As shown in Fig.\ref{fig:coordinate_system_trans}, the original $O-xyz$ Cartesian coordinate system $\mathcal{C}_0$ is shown using black lines. To establish the mapping between $\mathcal{C}_0$ and the above prolate spheroidal coordinate system, $\mathcal{C}_0$ needs to be translated and rotated at first. We define the Cartesian coordinate system after translation and rotation as $\mathcal{C}_1$, and the corresponding prolate spheroidal coordinate system as $\mathcal{C} _2$. $\mathcal{C}_1$ and $\mathcal{C} _2$ are illustrated using blue lines and red curves, respectively. The detailed process of the coordinate system transformation is described as follows.

\textbf{Coordinate system transformation from $\mathcal{C} _0$ to $\mathcal{C} _1$:} As illustrated in Fig.\ref{fig:coordinate_c0_to_c1}, we define the centre point of the line segment $\overline{\mathrm{P}_{\mathrm{TX}}\mathrm{P}_{\mathrm{DT}}}$, i.e., $O_1$, as the origin of $\mathcal{C} _1$, where the coordinates of $O_1$ in $\mathcal{C} _0$ are $\left( x_{\mathrm{c}},y_{\mathrm{c}},z_{\mathrm{c}} \right) $. To transform $\mathcal{C} _0$ into $\mathcal{C} _1$, we first translate $\mathcal{C} _0$ from the origin $O$ to the new origin $O_1$, and then rotate it clockwise through an angle of $\gamma_\mathrm{c}$ around the translated $x$-axis, where $\gamma_\mathrm{c}$ is the angle between the positive half of the $z$-axis and the vector $\overrightarrow{\mathrm{P}_{\mathrm{TX}}\mathrm{P}_{\mathrm{DT}}}$, which is given by
\begin{equation}
   \gamma _{\mathrm{c}}=\mathrm{arctan}\left( \frac{l_{\mathrm{DT}}\sin \gamma}{l_{\mathrm{DT}}\cos \gamma -l_{\mathrm{TX}}} \right) \in (0,\frac{\pi}{2}).
   \label{gamma_c_calculation}
\end{equation}
Hence, the coordinates $(x,y,z)$ in $\mathcal{C} _0$ are related to the coordinates $\left( x_1,y_1,z_1 \right) $ in $\mathcal{C}_1$ via $\mathbf{p}_1=\mathbf{R}_x\left( \gamma _{\mathrm{c}} \right) \left( \mathbf{p}-\mathbf{p}_{\mathrm{c}} \right) $, where $\mathbf{p}=\left[ x,y,z \right]^{\mathrm{T}}$ and $\mathbf{p}_1=\left[ x_1,y_1,z_1 \right]^{\mathrm{T}}$ are the vectorized representation for the same point in $\mathcal{C} _0$ and $\mathcal{C} _1$, respectively. In addition, $\mathbf{p}_{\mathrm{c}}=\left[ x_{\mathrm{c}},y_{\mathrm{c}},z_{\mathrm{c}} \right] ^{\mathrm{T}}=\left[ 0,\left( l_{\mathrm{DT}}\sin \gamma \right) /2,\left( l_{\mathrm{TX}}+l_{\mathrm{DT}}\cos \gamma \right) /2 \right] ^{\mathrm{T}}$ is the vectorized repsentation for the point $O_1$ $\left( x_{\mathrm{c}},y_{\mathrm{c}},z_{\mathrm{c}} \right)$ in $\mathcal{C}_0$, and $\mathbf{R}_x\left( \gamma _{\mathrm{c}} \right) $ is the rotation matrix of the coordinate system through angle $\gamma_\mathrm{c}$ around the translated $x$-axis. 

\textbf{Coordinate system transformation from $\mathcal{C} _1$ to $\mathcal{C} _2$:} We perform the transformation of the coordinate system from $\mathcal{C}_1$ to the prolate spheroidal coordinate system $\mathcal{C}_2$ and establish the mapping among $\mathcal{C} _0$, $\mathcal{C} _1$, and $\mathcal{C} _2$. Let the lengths of $\overline{O_1 \mathrm{P}_{\mathrm{DT}}}$ and $\overline{O_1 \mathrm{P}_{\mathrm{TX}}}$ be $a$, and the coordinates of $\mathrm{P}_{\mathrm{\mathrm{TX}}}$ and $\mathrm{P}_{\mathrm{\mathrm{DT}}}$ can be expressed as $(0,0,-a)$ and $(0,0,a)$ in $\mathcal{C} _1$. Let $O_1$ be the origin of $\mathcal{C}_2$, and the points $\mathrm{P}_{\mathrm{\mathrm{TX}}}$ and $\mathrm{P}_{\mathrm{\mathrm{DT}}}$ be the two foci of the prolate spheroidal system $\mathcal{C}_2$.\footnote{The establishment of $\mathcal{C}_2$ is not degenerate. There is always a unique and reversible correspondence between $\mathcal{C}_1$ and $\mathcal{C}_2$.} Therefore, the mapping among these three coordinates can be given by 
\begin{align}
\left[ \begin{array}{c}
	x_1\\
	y_1\\
	z_1\\
\end{array} \right] &=\left[ \begin{array}{c}
	x-x_{\mathrm{c}}\\
	\cos \gamma _{\mathrm{c}}\left( y-y_{\mathrm{c}} \right) -\sin \gamma _{\mathrm{c}}\left( z-z_{\mathrm{c}} \right)\\
	\sin \gamma _{\mathrm{c}}\left( y-y_{\mathrm{c}} \right) +\cos \gamma _{\mathrm{c}}\left( z-z_{\mathrm{c}} \right)\\
\end{array} \right] \notag
\\
&=\left[ \begin{array}{c}
	a\sqrt{\left( \sigma ^2-1 \right) \left( 1-\tau ^2 \right)}\cos \varphi\\
	a\sqrt{\left( \sigma ^2-1 \right) \left( 1-\tau ^2 \right)}\sin \varphi\\
	a\sigma \tau\\
\end{array} \right] ,
\label{equ:relation_C0_C1_C2}
\end{align}
where we use $(\sigma,\tau,\varphi)$ to express the prolate spheroidal coordinate system $\mathcal{C} _2$, $\sigma \in [1,\infty)$, $\tau \in [-1,1]$, and $\varphi \in [0,2\pi]$. The sum of distances between any point on the spheroid with given $\sigma$ and its two foci is $2a\sigma$.

Solving the equation (\ref{equ:relation_C0_C1_C2}), we have the set of equations (\ref{equ:relation_C0_C2}) that precisely constructs the mapping between $\mathcal{C}_0$ and $\mathcal{C}_2$. 

\begin{figure*}[b]
\begin{equation}\label{equ:relation_C0_C2}
 \begin{gathered}
  \begin{aligned}
    \begin{cases}
	x&=x_c+a\,\cos \left( \phi \right) \,\sqrt{-\left( \sigma ^2-1 \right) \,\left( \tau ^2-1 \right)}\\
	y&=y_c+a\,\sigma \,\tau \,\sin \left( \gamma _c \right) +a\,\cos \left( \gamma _c \right) \,\sin \left( \phi \right) \,\sqrt{-\left( \sigma ^2-1 \right) \,\left( \tau ^2-1 \right)}\\
	z&=z_c+a\,\sigma \,\tau \,\cos \left( \gamma _c \right) -a\,\sin \left( \gamma _c \right) \,\sin \left( \phi \right) \,\sqrt{-\left( \sigma ^2-1 \right) \,\left( \tau ^2-1 \right)}
    \end{cases}.
 \end{aligned}
 \end{gathered}
\end{equation}
\hrulefill
\end{figure*}

\begin{figure}[t] % Use figure* for full-width figure across two columns
\centering
\begin{subfigure}[t]{0.7\linewidth}
    \centering
    \includegraphics[width=\linewidth]{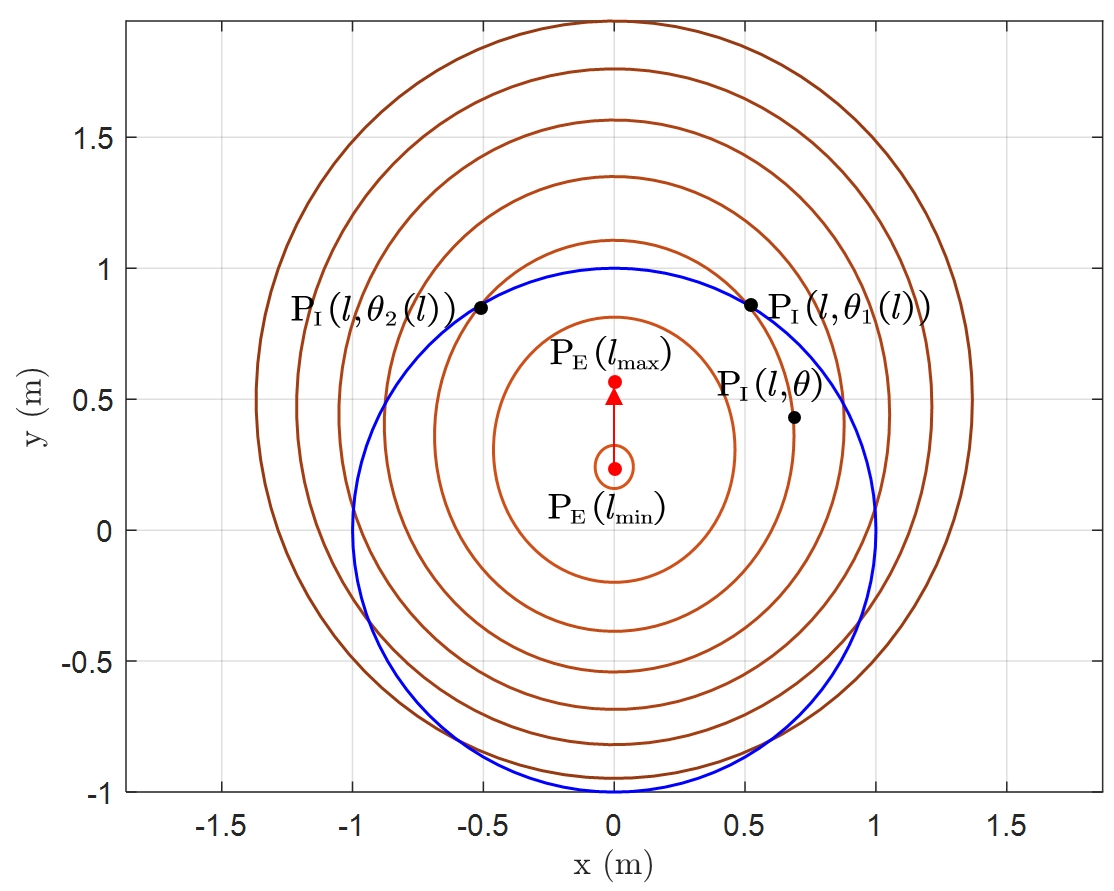}
    \caption{Point $\mathrm{P}_{\mathrm{E}}(l_{\min})$ is inside the RIS circle.}
    \label{fig:integral_range_a}
\end{subfigure}\hfil
\begin{subfigure}[t]{0.7\linewidth}
    \centering
    \includegraphics[width=\linewidth]{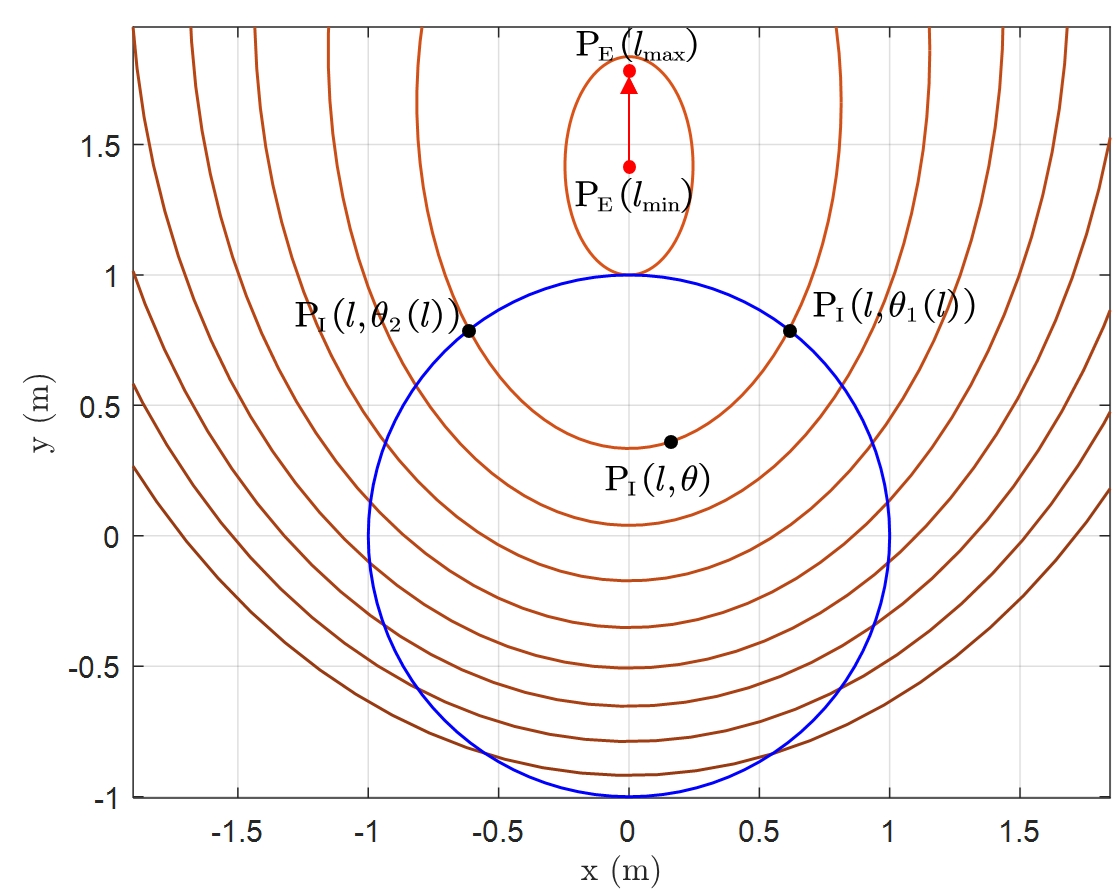}
    \caption{Point $\mathrm{P}_{\mathrm{E}}(l_{\min})$ is outside the RIS circle.}
    \label{fig:integral_range_b}
\end{subfigure}\hfil
\caption{Illustration of the intersection and the RIS circle.}
\label{fig:integarl_range_explain}
\end{figure}

Back to the expression (\ref{tilde_g_a_f_without_eta}), to get the additional phase shift $\phi_{\mathrm{des}}(x,y)$, we aim to simplify (\ref{tilde_g_a_f_without_eta}) by deriving the integral which is only related to the sum of distances $2a\sigma$. We still use $l$ to represent the sum of distances and let $l=2a\sigma\in[l_{\min},l_{\max})$, where $l_{\mathrm{min}}$ and $l_{\mathrm{max}}$ are respectively the minimal and the maximal sum of distances corresponding to the non-empty intersection $\mathcal{S} _{\mathrm{RIS}}^{\mathrm{S}}(l)$. The values of $l_{\mathrm{min}}$ and $l_{\mathrm{max}}$ can be calculated by traversing all possible points (i.e., the position of RIS elements) on the RIS surface or be calculated using (\ref{equ:relation_C0_C2}) where $[x,y,z]=[-R,0,0]$ is corresponding to $l_{\mathrm{max}}$, and $l_{\mathrm{min}}$ is the smallest $l$ for which the equation (\ref{equ:relation_C0_C1_C2}) holds.

Then, the parameterized expression of the intersected ellipse $\mathcal{S}^{\mathrm{S}}_{z}(l)$ on the $z=0$ plane with given $l$ is
\begin{align}
    \begin{cases}
	x=x_{\mathrm{E}}\left( l \right) +a_{\mathrm{E}}\left( l \right) \cos \left( \theta \right)\\
	y=y_{\mathrm{E}}\left( l \right) +b_{\mathrm{E}}\left( l \right) \sin \left( \theta \right)
 \label{equs_x_y_l_theta_substitution}
\end{cases},
\end{align}
where $(x_{\mathrm{E}}(l),y_{\mathrm{E}}(l))$ are the 2D coordinates on the $x-O-y$ plane in $\mathcal{C}_0$, representing the center point of the ellipse $\mathrm{P}_{\mathrm{E}}(l)$, $a_{\mathrm{E}}\left( l \right)$ and $b_{\mathrm{E}}\left( l \right)$ are the lengths of the minor and major axis of the corresponding ellipse. The geometric relationship of the RIS circle and the ellipse/elliptical arc is shown in Fig.\ref{fig:integarl_range_explain}, where the blue curve represents the RIS circle, and the orange ellipses/elliptical arcs like annual rings represent $\mathcal{S}^{\mathrm{S}}_{z}(l)$ with different $l$'s. Let $\mathrm{P}_{\mathrm{I}}(l,\theta)$ represent the point on the intersection ellipse $\mathcal{S}^{\mathrm{S}}_{z}(l)$, and $\theta\in \left[ -\pi/2,3\pi/2 \right] $ is the angle from the positive semi-axis of the minor axis of the intersected ellipse, which has the same direction as the positive half of the x-axis, to the vector $\overrightarrow{\mathrm{P}_{\mathrm{E}}(l)\mathrm{P}_{\mathrm{I}}(l,\theta)}$. 
$\mathrm{P}_{\mathrm{I}}\left( l,\theta _1\left( l \right) \right)$ and $\mathrm{P}_{\mathrm{I}}\left( l,\theta _2\left( l \right) \right)$ represent the two intersection points of the ellipse with given $l$ and the RIS circle, where $\theta_1(l)\in [-\pi/2,\pi/2]$ and $\theta_2(l) \in [\pi/2,3\pi/2]$ are the lower bound and the upper bound. 

{\color{black}
Next, by substituting $x$ and $y$ by $l$ and $\theta$ using (\ref{equ:relation_C0_C2}) and $l=2a\sigma$, the expression (\ref{g_a_f_hat_a_l_FT_form}) can be rewritten as 
\begin{align}
\hat{g}_{\mathrm{a}}(\omega )&=\int_{\mathcal{S}}{\frac{e^{j\phi _{\mathrm{des}}(l)}e^{-jwl}|J\left( l ,\theta \right) |}{l_{\mathrm{TX}}\left( l,\theta \right) l_{\mathrm{DT}}\left( l,\theta \right)}\mathrm{d}l\mathrm{d}\theta}
\\
&=\int_{l_{\max}}^{l_{\max}}{\int_{\theta _1\left( l \right)}^{\theta _2\left( l \right)}{K\left( l,\theta \right) e^{j\phi _{\mathrm{des}}(l)}e^{-jwl}\mathrm{d}\theta}\mathrm{d}l},
\label{equation:g_a_w_S_integral_before_simplify}
\end{align}
where $K\left( l,\theta \right) =\frac{|J\left( l,\theta \right) |}{l_{\mathrm{TX}}\left( l,\theta \right) l_{\mathrm{DT}}\left( l,\theta \right)}$, and $|J\left( l,\theta \right) |=|\frac{\partial \left( x,y \right)}{\partial \left( l,\theta \right)}|$ is the absolute value of the determinant of the Jacobian matrix with given $l$. 

With the increase of $l$, the center point of the intersected ellipse $\mathrm{P}_{\mathrm{E}}(l)$ changes from $\mathrm{P}_{\mathrm{E}}(l_{\min})$ to $\mathrm{P}_{\mathrm{E}}(l_{\max})$.
Particularly, when $l=l_{\min}$, there is only one point in the intersection set $\mathcal{S} _{\mathrm{RIS}}^{\mathrm{S}}(l)$, i.e., $\mathrm{P}_{\mathrm{E}}(l_{\min})$ which is the point that the spheroid initially intersects with the $z=0$ plane. As shown in Fig.\ref{fig:integarl_range_explain}, The initial point $\mathrm{P}_{\mathrm{E}}(l_{\min})$ may fall inside the RIS circle (see Fig.\ref{fig:integral_range_a}) or outside the RIS circle (see Fig.\ref{fig:integral_range_b}). Different scenarios cause different integral ranges of the equation (\ref{equation:g_a_w_S_integral_before_simplify}). 
}

For the scenario where $\mathrm{P}_{\mathrm{E}}(l_{\min})$ is inside the RIS circle, the equation (\ref{equation:g_a_w_S_integral_before_simplify}) can be rewritten as
\begin{align}
&\hat{g}_{\mathrm{a}}(\omega )=\int\limits_{l_{\min}}^{l_{\mathrm{mid}}}{\left[ \int\limits_{-\frac{\pi}{2}}^{\frac{3}{2}\pi}{K\left( l,\theta \right) \mathrm{d}\theta} \right]}e^{j\phi _{\mathrm{des}}(l)}e^{-jwl}\mathrm{d}l \notag
\\
&+\int\limits_{l_{\min}}^{l_{\mathrm{mid}}}{\left[ \int\limits_{-\frac{\pi}{2}}^{\theta _1\left( l \right)}{K\left( l,\theta \right) \mathrm{d}\theta +\int\limits_{\theta _2\left( l \right)}^{\frac{3}{2}\pi}{K\left( l,\theta \right) \mathrm{d}\theta}} \right]}e^{j\phi _{\mathrm{des}}(l)}e^{-jwl}\mathrm{d}l
\\
&=\int_{-\infty}^{\infty}{A\left( l \right)}e^{j\phi _{\mathrm{des}}(l)}\mathbb{I} _{l_{\min}}^{l_{\max}}e^{-jwl}\mathrm{d}l,
\label{g_a_w_A_sigma_substitution_in_circle}
\end{align}
where $l_{\mathrm{mid}}>l_{\mathrm{min}}$ is the boundary value of $l$ when $\mathcal{S} _{\mathrm{RIS}}^{\mathrm{S}}(l)$ changes from the ellipse to the elliptical arc, and $A(l)$ is the amplitude modulation function given by
\begin{align}
A\left( l \right) =\begin{cases}
	\int\limits_{-\frac{\pi}{2}}^{\frac{3}{2}\pi}{K\left( l,\theta \right) \mathrm{d}\theta },\!\!\!\!&l\in \left[ l_{\min},l_{\mathrm{mid}} \right]\\
	\int\limits_{-\frac{\pi}{2}}^{\theta _1\left( l \right)}{K\left( l,\theta \right) \mathrm{d}\theta} +\int\limits_{\theta _2\left( l \right)}^{\frac{3}{2}\pi}{K\left( l,\theta \right) \mathrm{d}\theta},\!\!\!\! &l\in \left[ l_{\mathrm{mid}},l_{\max} \right]
 \label{equ:A_l_in_circle}
\end{cases}.
\end{align}
When $l_{\mathrm{mid}}<l<l_{\max}$, the solutions of  $\theta_1(l)$ and $\theta_2(l)$ can be numerically calculated by substituting (\ref{equs_x_y_l_theta_substitution}) into
\begin{align}
    \begin{cases}
	x^2+y^2=R^2\\
	z=0\\
\end{cases},
\label{equ:RIS_circle}
\end{align}
which represents the RIS circle. 

For the scenario where $\mathrm{P}_{\mathrm{E}}(l_{\min})$ is outside the RIS circle, the equation (\ref{equation:g_a_w_S_integral_before_simplify}) can be rewritten as
\begin{align}
\hat{g}_{\mathrm{a}}(\omega )&=\int_{l_{\min}}^{l_{\max}}{A\left( l \right) e^{j\phi _{\mathrm{des}}(l)}e^{-jwl}\mathrm{d}l}
\\
&=\int_{-\infty}^{\infty}{A\left( l \right)}e^{j\phi _{\mathrm{des}}(l)}\mathbb{I} _{l_{\min}}^{l_{\max}}e^{-jwl}\mathrm{d}l,
\label{g_a_w_A_sigma_substitution_out_circle}
\end{align}
where the amplitude modulation function $A(l)$ is given by 
\begin{align}
A\left( l \right) =\int_{-\frac{\pi}{2}}^{\theta _1\left( l \right)}\!\!\!{K\left( l,\theta \right) \mathrm{d}\theta +\int_{\theta _2\left( l \right)}^{\frac{3}{2}\pi}\!\!\!{K\left( l,\theta \right) \mathrm{d}\theta}},l\in \left[ l_{\min},l_{\max} \right] ,
\label{equ:A_l_out_circle}
\end{align}
where the solutions of  $\theta_1(l)$ and $\theta_2(l)$ can be numerically calculated by substituting (\ref{equs_x_y_l_theta_substitution}) into the equations (\ref{equ:RIS_circle}) similarly. 

Finally, since (\ref{g_a_w_A_sigma_substitution_in_circle}) and (\ref{g_a_w_A_sigma_substitution_out_circle}) have the same structures as (\ref{g_a_f_hat_a_l_FT_form}), the SPM can be used to design $\phi_{\mathrm{des}}(l)$, making $|\tilde{g}_{\mathrm{a}}(\omega )|^2$ large and flat over the desired bandwidth. The procedure of the proposed SFFT-based RIS wideband beamforming design is summarized in Algorithm \ref{alg:RIS_WBF}.

\begin{algorithm}[htbp]
\caption{The SFFT-based RIS Wideband beamforming Design}
\label{alg:RIS_WBF}
\begin{algorithmic}[1]
\REQUIRE $R$, $\Delta$, $\mathcal{S}_{\mathrm{D}}$, $l_{\mathrm{TX}}$, $l_{\mathrm{DT}}$, $\gamma_{\mathrm{c}}$, $\phi_{\mathrm{std}}(x,y)$.
\STATE Establish the mapping between $\mathcal{C}_0$ and $\mathcal{C}_2$ via (\ref{equ:relation_C0_C2}).
\STATE Let $z=0$ in (\ref{equ:relation_C0_C2}) and get the parametric equations (\ref{equs_x_y_l_theta_substitution}).
\STATE Calculate the minimal and the maximal sum of distances $l_{\mathrm{min}}$ and $l_{\mathrm{max}}$ by traversing all possible points in $\mathcal{S}_{\mathrm{D}}$.
\STATE Evenly sample values from the continuous interval $[l_{\mathrm{min}},l_{\mathrm{max}}]$, producing a discrete set $\mathcal{L}$.
\FOR {each element $l\in\mathcal{L}$}
\STATE Calculate the corresponding $\theta_1(l)$ and $\theta_2(l)$ numerically by substituting (\ref{equs_x_y_l_theta_substitution}) into (\ref{equ:RIS_circle}). 
\IF {the initial intersection point $P_{\mathrm{E}}(l_{\mathrm{min}})$ is inside the RIS circle}
\STATE Calculate $A(l)$ via (\ref{equ:A_l_in_circle}).
\ELSE
\STATE Calculate $A(l)$ via (\ref{equ:A_l_out_circle}).
\ENDIF
\ENDFOR
\STATE Calculate $\phi_{\mathrm{des}}'(l)$ from $A(l)$ using SPM via (\ref{phi_d}).
\STATE Calculate $\phi_{\mathrm{des}}(l)$ from $\phi_{\mathrm{des}}'(l)$ using SPM via (\ref{phi_solution}).
\STATE Map $\phi_{\mathrm{des}}(l)$ to $\phi_{\mathrm{des}}(x,y)$ via (\ref{equ:phi_x_y_phi_l}). 
\STATE Calculate the phase shift $\phi(x,y)$ according to (\ref{phi_added}).
\ENSURE $\phi(x,y)$.
\end{algorithmic}
\end{algorithm}

\begin{figure*}[ht] % Use figure* for full-width figure across two columns
\centering
\begin{subfigure}[t]{0.28\linewidth}
    \centering
    \includegraphics[width=\linewidth,height=\linewidth]{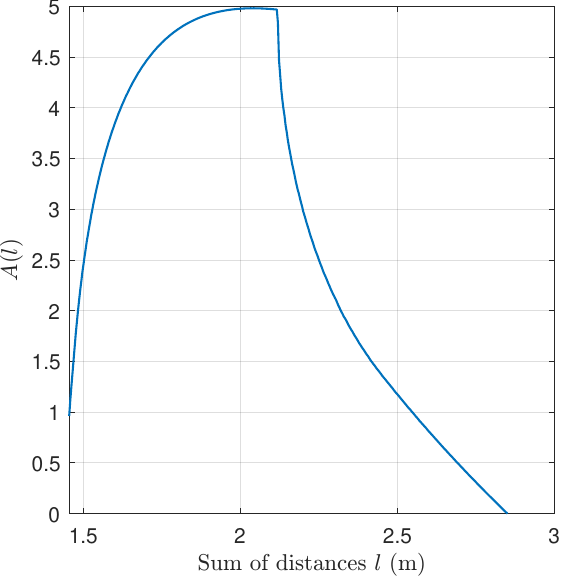}
    \caption{Modulation function $A(l)$.}
    \label{fig:amplitude-factor_non_boresight}
\end{subfigure}\hfil
\begin{subfigure}[t]{0.28\linewidth}
    \centering
    \includegraphics[width=\linewidth,height=\linewidth]{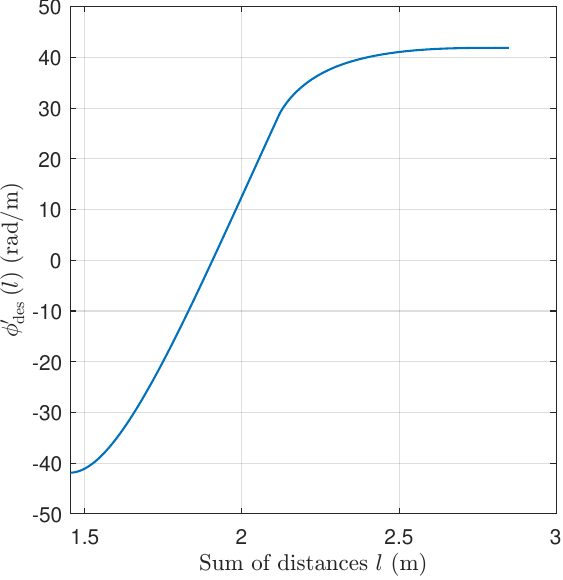}
    \caption{Instantaneous frequency $\phi_{\mathrm{des}} ' (l)$.}
    \label{fig:phi-prime-l_non_boresight}
\end{subfigure}\hfil
\begin{subfigure}[t]{0.28\linewidth}
    \centering
    \includegraphics[width=\linewidth,height=\linewidth]{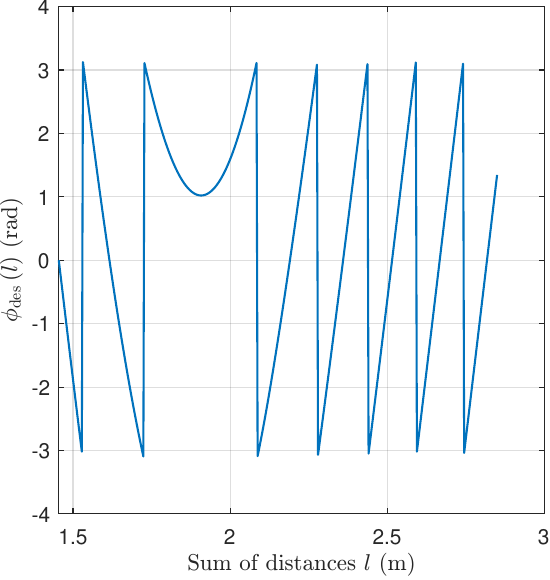}
    \caption{Additional phase shift $\phi_{\mathrm{des}} (l)$.}
    \label{fig:phi-l_non_boresight}
\end{subfigure}

\caption{Results of SPM when the location of the DT is arbitrary.}
\label{fig:results of boresight SPM illustration non-boresight}
\end{figure*}

\begin{figure*}[ht] % Use figure* for full-width figure across two columns
\centering
\begin{subfigure}[t]{0.33\linewidth}
    \centering
    \includegraphics[width=\linewidth]{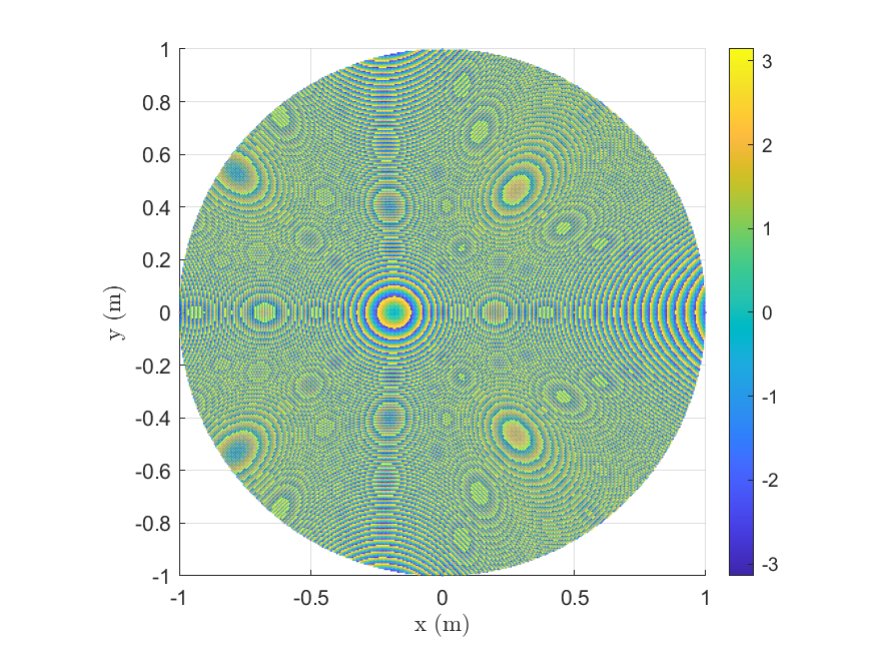}
    \caption{$\phi_{\mathrm{std}}(x,y)$.}
    \label{fig:phase_shift_non_boresight_a}
\end{subfigure}\hfil
\begin{subfigure}[t]{0.33\linewidth}
    \centering
    \includegraphics[width=\linewidth]{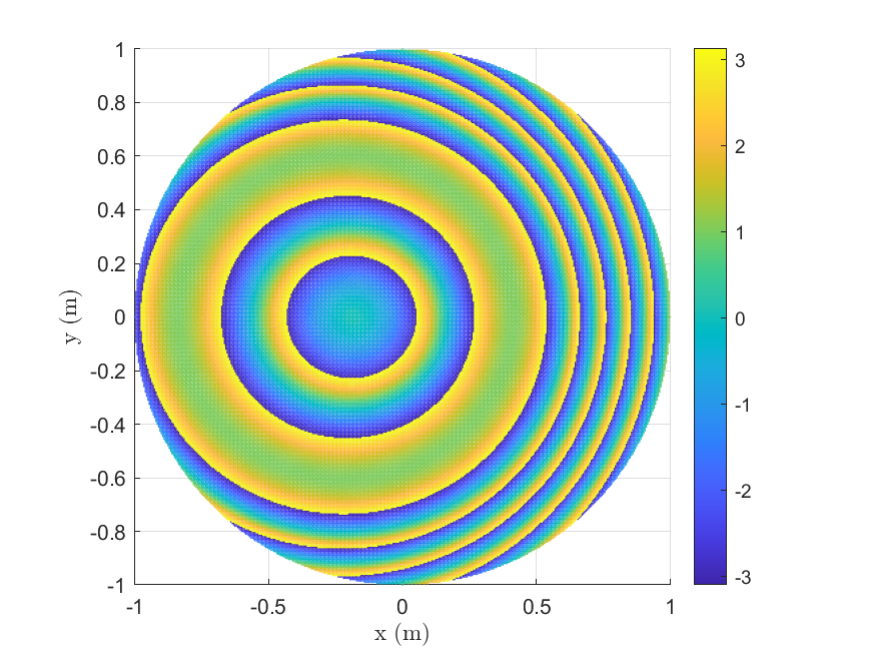}
    \caption{$\phi_{\mathrm{des}}(x,y)$.}
    \label{fig:phase_shift_non_boresight_b}
\end{subfigure}\hfil
\begin{subfigure}[t]{0.33\linewidth}
    \centering
    \includegraphics[width=\linewidth]{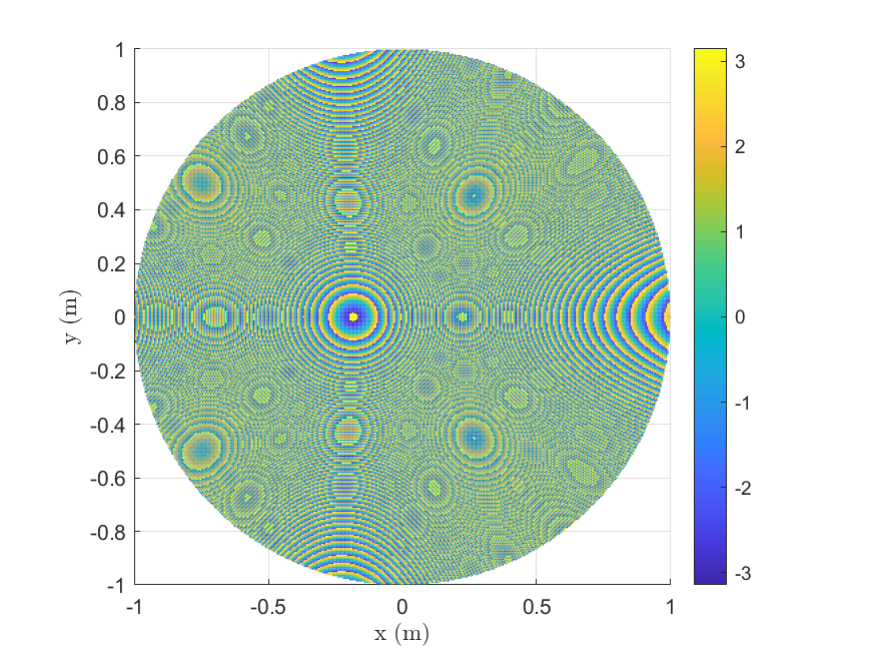}
    \caption{$\phi(x,y)=\phi_{\mathrm{std}}(x,y)+\phi_{\mathrm{des}}(x,y)$.}
    \label{fig:phase_shift_non_boresight_c}
\end{subfigure}

\caption{Phase shifts when the location of the DT is arbitrary.}
\label{fig:results of Non-boresight phase shift}
\end{figure*}

\begin{figure*}[t] % Use figure* for full-width figure across two columns
\centering
\begin{subfigure}[t]{0.28\linewidth}
    \centering
    \includegraphics[width=\linewidth]{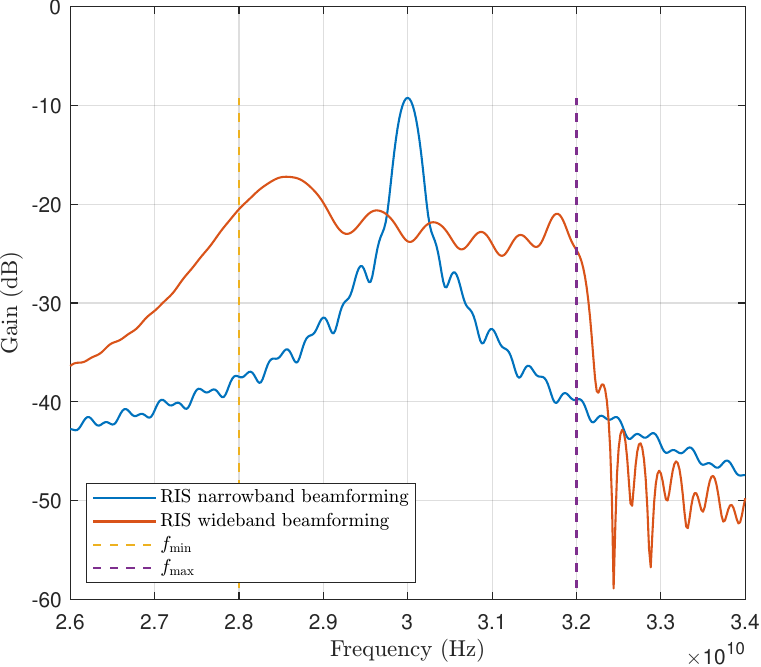}
    \caption{$l_{\mathrm{DT}}=1\mathrm{m},\gamma_{\mathrm{c}}=30^{\circ}$.}
    \label{fig:beampattern_non_boresight_a}
\end{subfigure}\hfil
\begin{subfigure}[t]{0.28\linewidth}
    \centering
    \includegraphics[width=\linewidth]{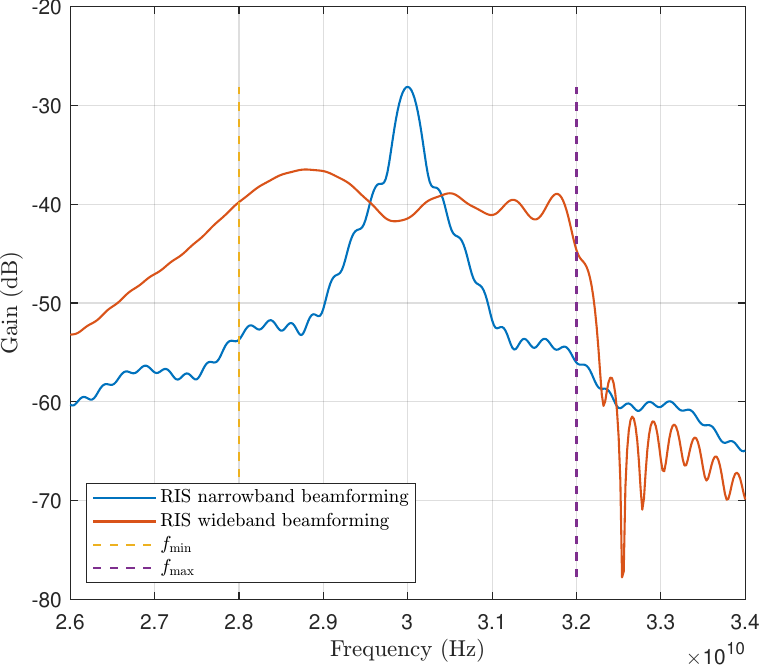}
    \caption{$l_{\mathrm{DT}}=10\mathrm{m},\gamma_{\mathrm{c}}=30^{\circ}$.}
    \label{fig:beampattern_non_boresight_b}
\end{subfigure}\hfil
\begin{subfigure}[t]{0.28\linewidth}
    \centering
    \includegraphics[width=\linewidth]{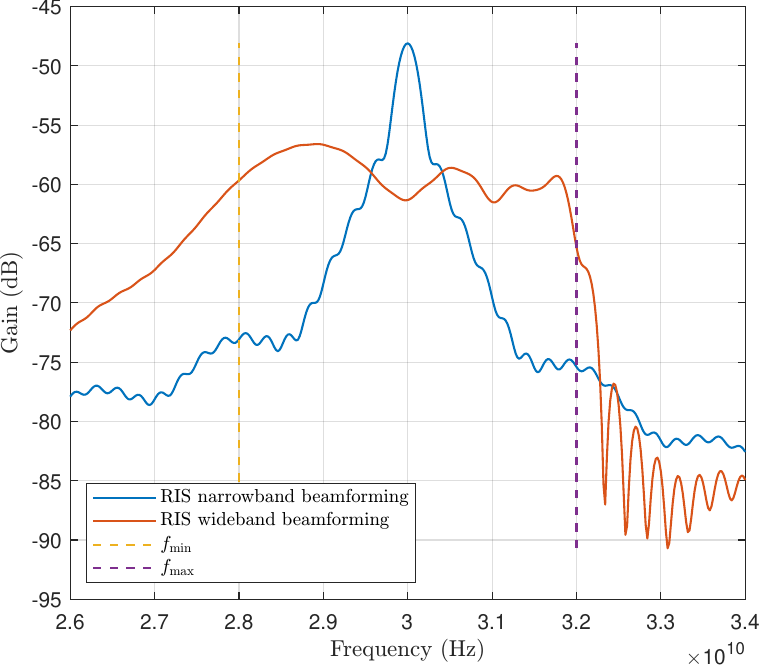}
    \caption{$l_{\mathrm{DT}}=100\mathrm{m},\gamma_{\mathrm{c}}=30^{\circ}$.}
    \label{fig:beampattern_non_boresight_c}
\end{subfigure}

\begin{subfigure}[t]{0.28\linewidth}
    \centering
    \includegraphics[width=\linewidth]{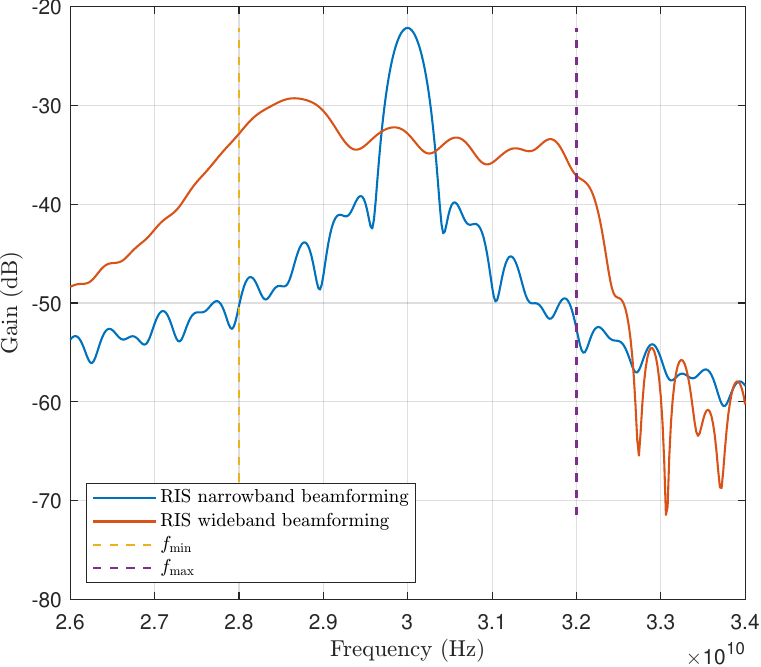}
    \caption{$l_{\mathrm{DT}}=5\mathrm{m},\gamma_{\mathrm{c}}=10^{\circ}$.}
    \label{fig:beampattern_non_boresight_d}
\end{subfigure}\hfil
\begin{subfigure}[t]{0.28\linewidth}
    \centering
    \includegraphics[width=\linewidth]{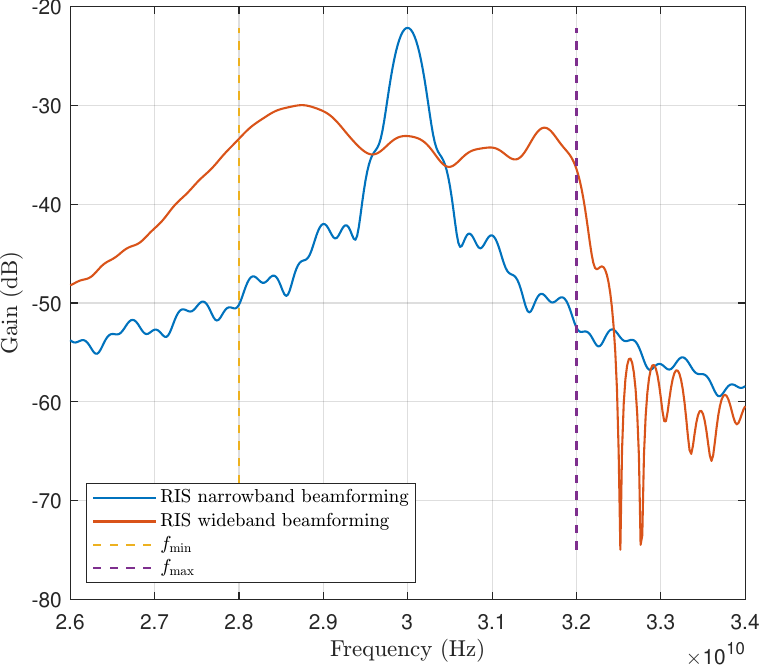}
    \caption{$l_{\mathrm{DT}}=5\mathrm{m},\gamma_{\mathrm{c}}=20^{\circ}$.}
    \label{fig:beampattern_non_boresight_e}
\end{subfigure}\hfil
\begin{subfigure}[t]{0.28\linewidth}
    \centering
    \includegraphics[width=\linewidth]{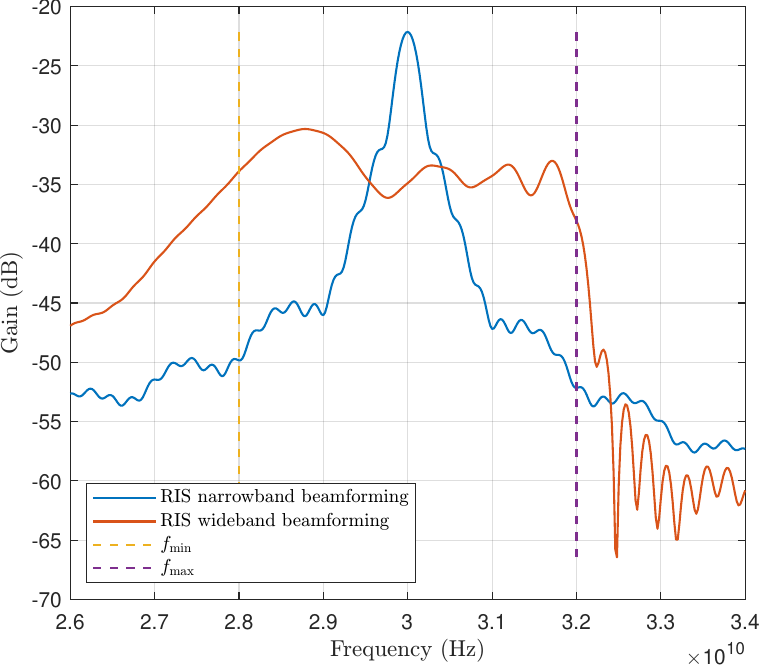}
    \caption{$l_{\mathrm{DT}}=5\mathrm{m},\gamma_{\mathrm{c}}=30^{\circ}$.}
    \label{fig:beampattern_non_boresight_f}
\end{subfigure}
\caption{Beampatterns when the location of the DT is arbitrary.}
\label{fig:results of non-boresight beampattern}
\end{figure*}

Next, to show the effectiveness of the proposed method in the non-boresight scenario, we consider a wideband system setup with parameters as follows: $f_{\mathrm{c}}=30$ GHz, $B=4$ GHz, $R=1$ m, $l_{\mathrm{TX}}=0.5$ m, $l_{\mathrm{DT}}=1$ m, and $\gamma_{\mathrm{c}}=30^\circ$. The results of SPM are illustrated in Fig. \ref{fig:results of boresight SPM illustration non-boresight}. The corresponding phase shifts are shown in Fig.\ref{fig:results of Non-boresight phase shift}. Furthermore, in Fig.\ref{fig:results of non-boresight beampattern}, we show the beampatterns with different $l_{\mathrm{DT}}$ and $\gamma_{\mathrm{c}}$. We can see that the proposed wideband beamforming achieves a relatively large and flat gain across the desired frequency range in all configurations. A more detailed analysis is provided in Section.\ref{sec:simulation results}.

\textbf{Extension of the proposed method:} The proposed SFFT-based method can be easily extended to the general case with any RIS's shape, TX's location, and expected frequency-domain beampattern. For any shape of the RIS, one only needs to change the range of the integral in (\ref{equation:g_a_w_S_integral_before_simplify}). For any location of the TX, one can still derive the relationship like (\ref{equ:relation_C0_C1_C2}), solve the problem similarly by constructing the equations of the intersection line, and get the approximated solution using SPM. For an arbitrary wideband beampattern, one simply needs to match the RIS phase shift with the amplitude modulation function according to (\ref{phi_dd}). The adaptation of the proposed method via space-frequency transformation shows great potential in RIS-assisted wideband applications.

\vspace{-1 mm}
\section{Simulation Results}
\label{sec:simulation results}
In this section, we provide the simulation results to demonstrate the effectiveness of the proposed SFFT-based RIS wideband beamforming, and its benefits in improving both communication rate and sensing resolution. Unless otherwise specified, the following setup is used: $f_{\mathrm{c}}=30$ GHz, $B=4$ GHz, $R=1$ m, $l_{\mathrm{TX}}=0.5$ m, $l_{\mathrm{DT}}=5$ m, and $\gamma_{\mathrm{c}}=10^\circ$, $f_{\min}=28$ GHz, $f_{\max}=32$ GHz. \textcolor{black}{Under such conditions, $N_{\mathrm{RIS}} \approx 1.2 \times 10^6$.}

\subsection{Beampattern Analysis}
\label{numerical_results_beampattern}

\begin{figure*}[t] % Use figure* for full-width figure across two columns
\centering
\begin{subfigure}[t]{0.28\linewidth}
    \centering
    \includegraphics[width=\linewidth]{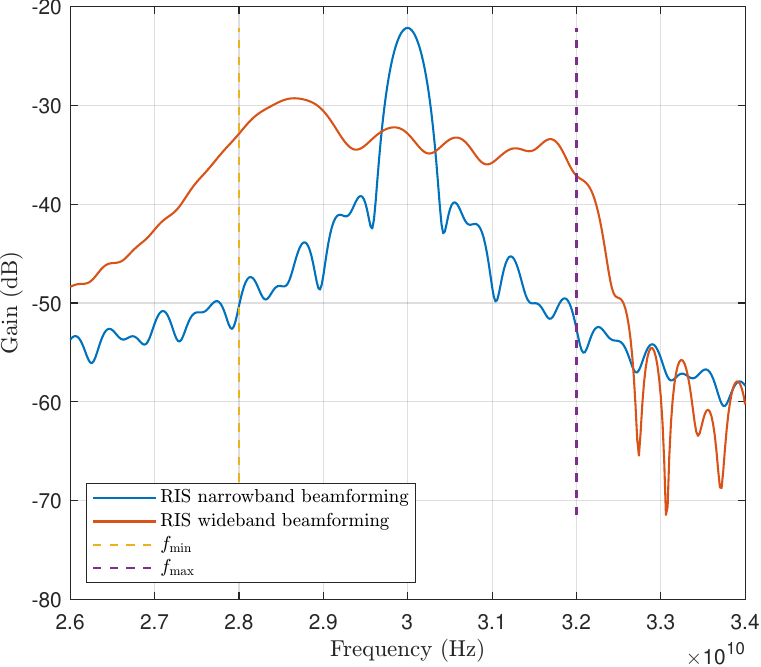}
    \caption{$R=1$ m.}
    \label{fig:results_beampattern_a}
\end{subfigure}\hfil
\begin{subfigure}[t]{0.28\linewidth}
    \centering
    \includegraphics[width=\linewidth]{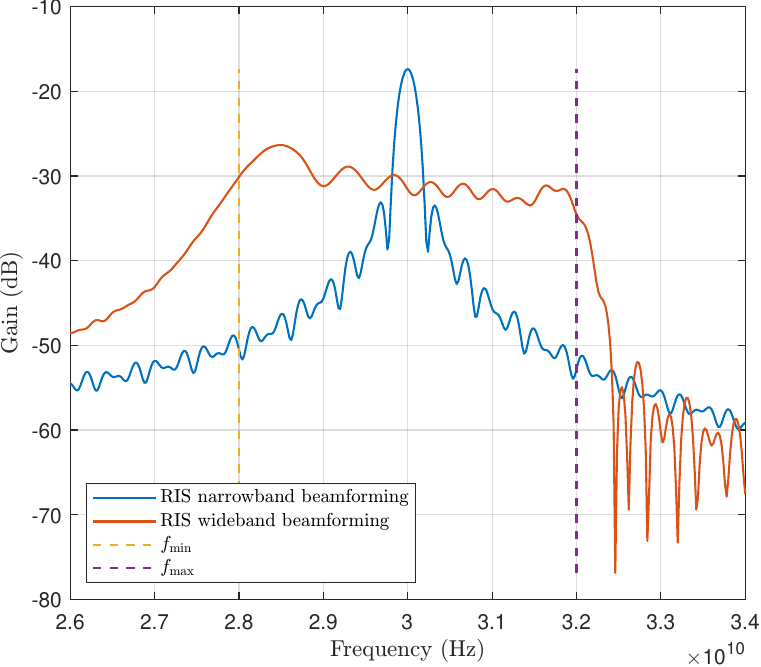}
    \caption{$R=1.5$ m.}
    \label{fig:results_beampattern_b}
\end{subfigure}\hfil
\begin{subfigure}[t]{0.28\linewidth}
    \centering
    \includegraphics[width=\linewidth]{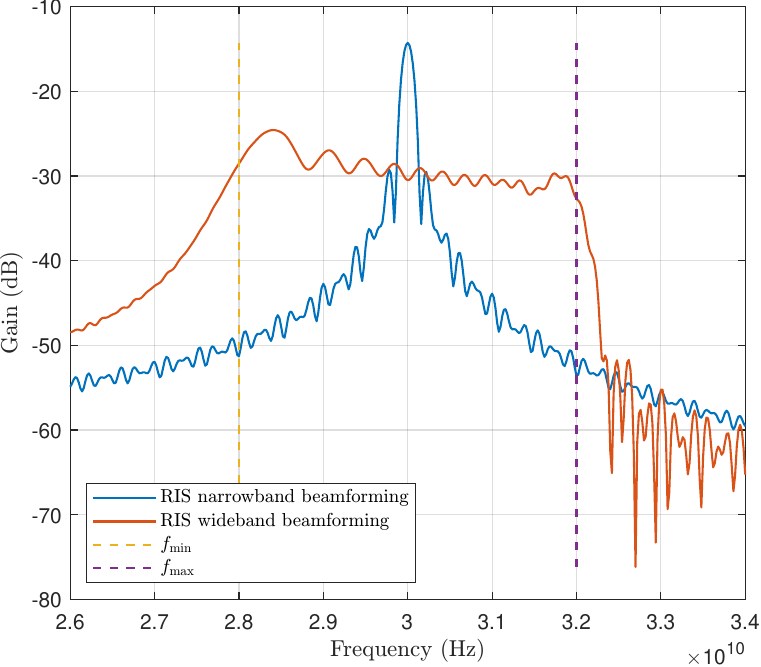}
    \caption{$R=2$ m.}
    \label{fig:results_beampattern_c}
\end{subfigure}

\caption{Beampatterns with different sizes of the RIS.}
\label{fig:results_part_beampattern_size}
\end{figure*}

Fig.\ref{fig:results_part_beampattern_size} illustrates the frequency-domain beampattern steering to the DT over a wide frequency band with different sizes of the RIS. As shown in Fig.\ref{fig:results_part_beampattern_size}, the gains of the proposed RIS wideband beamforming are large and flat over the desired 4 GHz bandwidth. The difference in the gains over the desired bandwidth is less than 6 dB in all three configurations. On the contrary, the gains of the RIS narrowband beamforming are only concentrated around the $f_{\mathrm{c}}$ and drop quickly in the region away from $f_{\mathrm{c}}$. Moreover, as $R$ increases, the advantages of the proposed RIS wideband beamforming over narrowband beamforming become more significant. The main lobe of the beampattern for narrowband beamforming becomes narrower, while the beampattern for the proposed RIS wideband beamforming exhibits enhanced flatness over the desired frequency band and a steeper roll-off characteristic beyond the desired frequency band.

\begin{figure*}[t] % Use figure* for full-width figure across two columns
\centering
\begin{subfigure}[t]{0.28\linewidth}
    \centering
    \includegraphics[width=\linewidth]{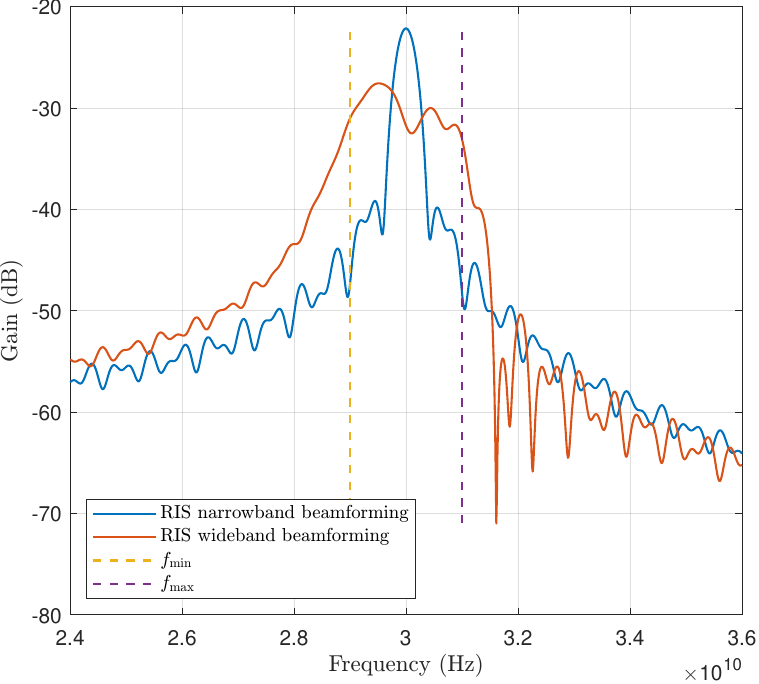}
    \caption{$B=2$ GHz.}
    \label{fig:results_beampattern_bandwidth_a}
\end{subfigure}\hfil
\begin{subfigure}[t]{0.28\linewidth}
    \centering
    \includegraphics[width=\linewidth]{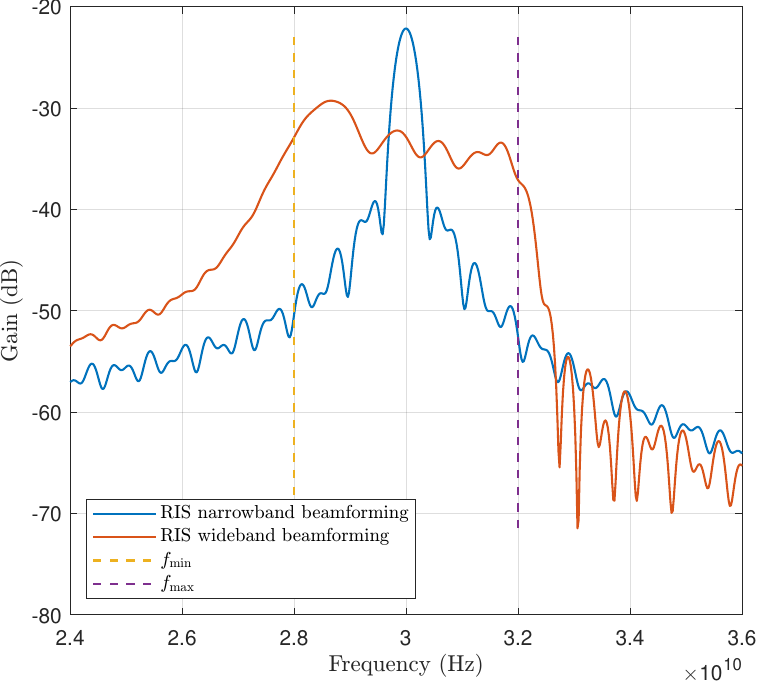}
    \caption{$B=4$ GHz.}
    \label{fig:results_beampattern_bandwidth_b}
\end{subfigure}\hfil
\begin{subfigure}[t]{0.28\linewidth}
    \centering
    \includegraphics[width=\linewidth]{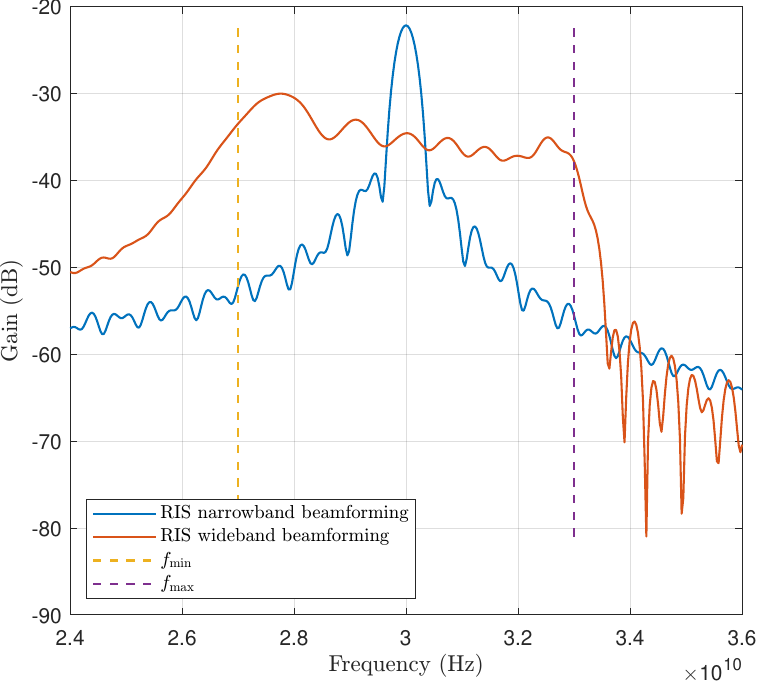}
    \caption{$B=6$ GHz.}
    \label{fig:results_beampattern_bandwidth_c}
\end{subfigure}

\caption{Beampatterns with different bandwidths.}
\label{fig:results_part_beampattern_bandwidth}
\end{figure*}
Fig.\ref{fig:results_part_beampattern_bandwidth} demonstrates the frequency-domain beampattern with different desired bandwidths. The proposed RIS wideband beamforming design is adaptable to a wide range of signal bandwidths. For the narrowband beamforming, as $B$ increases, the frequency-domain width of its main lobe remains unchanged, thereby making a limited effective bandwidth. On the contrary, even as $B$ increases from 2 GHz to 4 GHz, the gains of the proposed RIS wideband beamforming remain flat and large over the bandwidth, thereby making a wide effective bandwidth. When the desired bandwidth $B$ increases from 2 GHz to 6 GHz, the gains of the beampattern are properly and strictly limited within the desired bandwidth.

\subsection{Communication Performance}
\label{numerical_results_communication_performance}
\begin{figure}
    \centering
\begin{subfigure}[t]{0.8\linewidth}
    \centering
    \includegraphics[width=\linewidth]{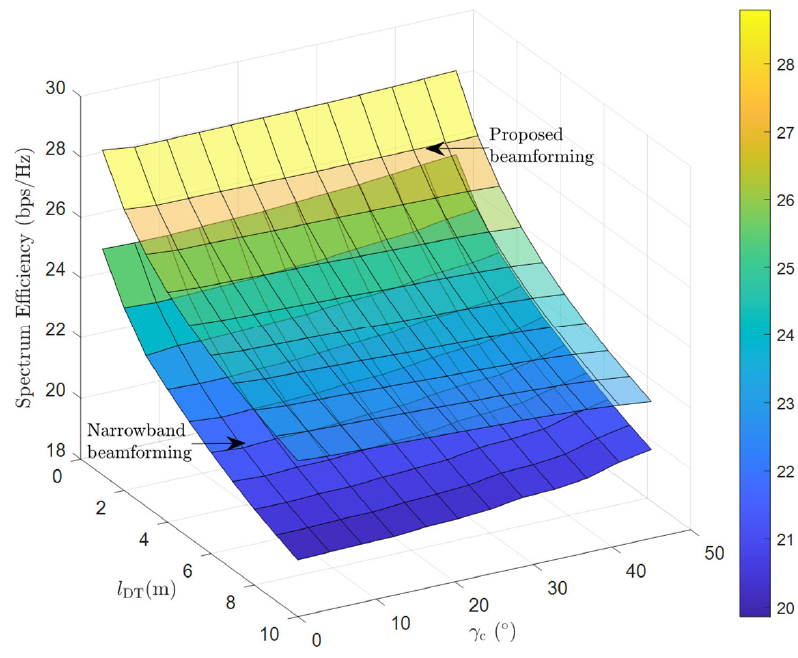}
    \caption{Comparison of the spectrum efficiency.}
    \label{fig:communication_compare_a}
\end{subfigure}\hfil
\begin{subfigure}[t]{0.7\linewidth}
    \centering
    \includegraphics[width=\linewidth]{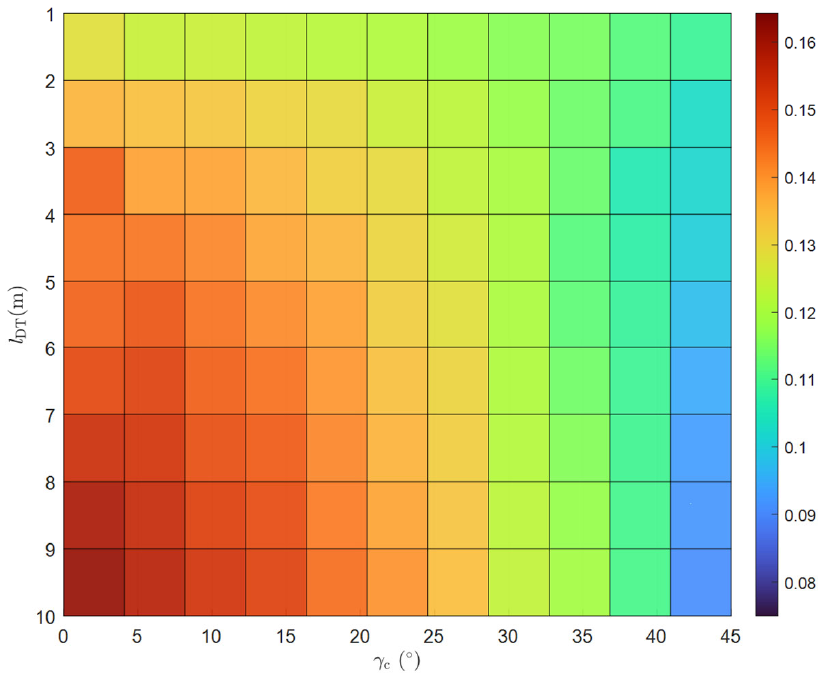}
    \caption{The increment of the spectrum efficiency.}
    \label{fig:communication_compare_b}
\end{subfigure}
    \caption{Performance comparison between the proposed RIS wideband beamforming and the standard narrowband beamforming in the communication rate.}
    \label{fig:communication_compare}
\end{figure}
\textcolor{black}{We illustrate the communication performance using the example of the wideband multicarrier communication system.}
The wideband channel over $[f_{\mathrm{c}}-B/2,f_{\mathrm{c}}+B/2]$ is first split into $N_{\mathrm{sub}}$ sub-bands. We define $\{f_k\}_{k=1}^{N_{\mathrm{sub}}}$ as $N_{\mathrm{sub}}$ equally spaced frequencies over $B$ and use $\eta_k$ to represent the power allocated over the $k$-th sub-band, where $\eta_k=\eta$, $\forall k =1,2,\cdots,N_{\mathrm{sub}}$. The bandwidth of each sub-band is $\Delta_{\mathrm{sub}} =B/ N_{\mathrm{sub}}$. We use the frequency selective thermal noise model and define $\mathrm{n}(f)$ as the noise power spectral density which can be calculated via $\mathrm{n}(f)=\frac{\hbar f}{\mathrm{exp}(\frac{\hbar f}{k_{\mathrm{btz}}T})-1}$, where $\hbar $ is the Planck constant, $k_{\mathrm{btz}}$ is the Boltzmann constant, and $T$ is the system temperature. The spectrum efficiency corresponding to the wideband SISO system is expressed as $R=\frac{1}{N_{\mathrm{sub}}}\sum_{k=1}^{N_{\mathrm{sub}}}{\log _2}\left( 1+\frac{\eta |g(f_k)|^2}{\mathrm{n(}f_k)\Delta _{\mathrm{sub}}} \right).$ We adopt the following simulation setups: $l_{\mathrm{DT}}\in [1,10]$ m, $\gamma_{\mathrm{c}}\in [0^\circ,45^\circ]$, $N_{\mathrm{sub}}=200$, $T=290$ Kelvin, $\hbar =6.625\times10^{-34}$ Joule$\cdotp$sec,  $k_{\mathrm{btz}}=1.3806\times10^{-23}$ Joule/Kelvin, and $\eta=5$ mW.

As shown in Fig.\ref{fig:communication_compare}, the spectrum efficiency between the proposed RIS wideband beamforming and the standard narrowband beamforming is compared. The proposed wideband beamforming makes the gain large and flat over the desired bandwidth, enabling wider effective bandwidth. Therefore, the spectrum efficiency of the proposed beamforming is always better than the standard narrowband beamforming for all the configurations of the DT's location. Specifically, Fig.\ref{fig:communication_compare_b} demonstrates that the proposed SFFT-based RIS wideband beamforming outperforms the standard narrowband beamforming by achieving an increment ranging from 8\% to 16\% under the given parameters.

\subsection{Sensing Performance}
\label{numerical_results_sensing_performance}

\begin{figure*}[t] % Use figure* for full-width figure across two columns
\centering
\begin{subfigure}[t]{0.3\linewidth}
    \centering
    \includegraphics[width=\linewidth]{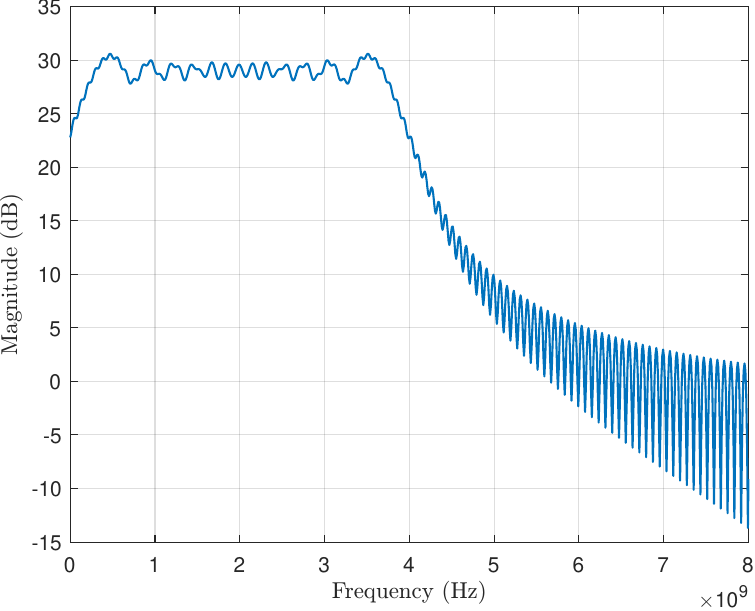}
    \caption{The spectrum of the LFM signal.}
    \label{fig:results_SP_a}
\end{subfigure}\hfil
\begin{subfigure}[t]{0.3\linewidth}
    \centering
    \includegraphics[width=\linewidth]{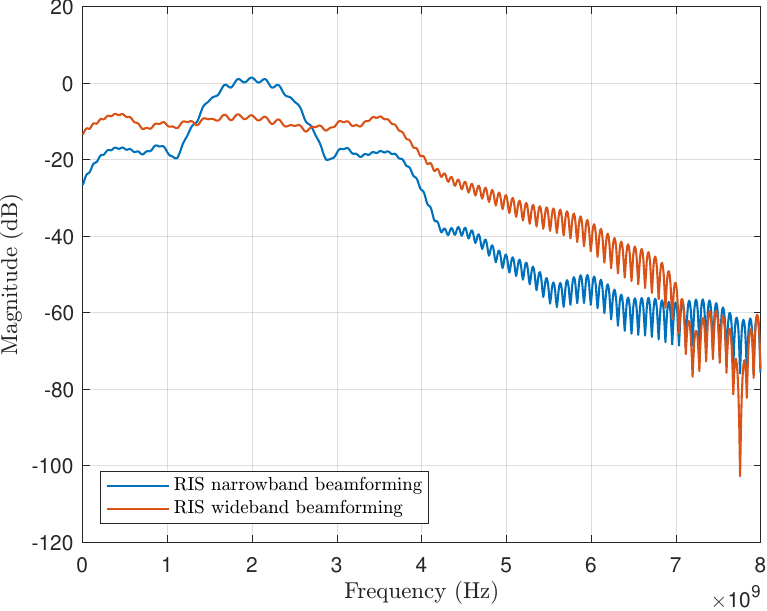}
    \caption{The spectrum after beamforming.}
    \label{fig:results_SP_b}
\end{subfigure}\hfil
\begin{subfigure}[t]{0.3\linewidth}
    \centering
    \includegraphics[width=\linewidth]{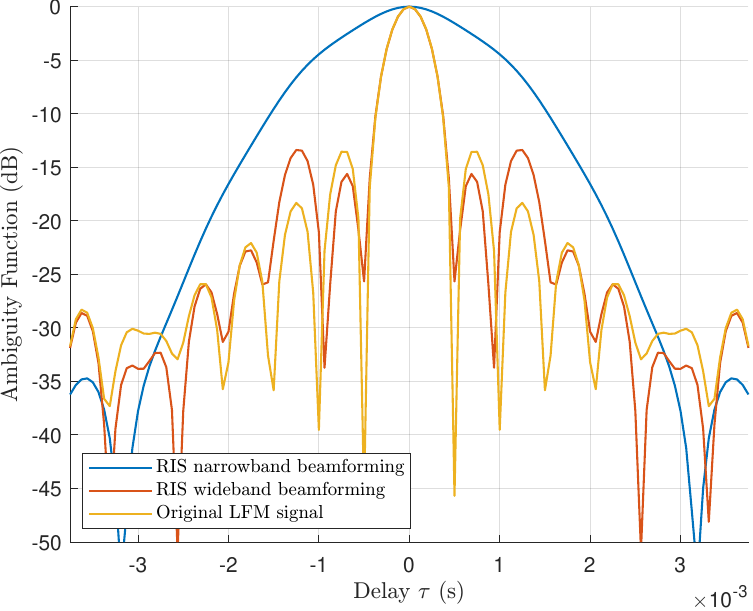}
    \caption{Ambiguity function Comparision.}
    \label{fig:results_SP_c}
\end{subfigure}

\caption{Sensing performance comparison ($l_{\mathrm{DT}}=10$ m).}
\label{fig:results_sensing_performance}
\end{figure*}
We compare the distance resolution between the proposed RIS wideband beamforming and the standard narrowband beamforming, taking the wideband linear frequency modulation (LFM) radar signal as an example. Theoretically, the relationship between the distance resolution $\Delta D$ and the signal bandwidth $B$ is given by $\Delta D=\frac{c}{2B}$. However, narrowband beamforming may limit the effective bandwidth of the signal, which will deteriorate the $\Delta D$. To analyze it, we use the ambiguity function as the metric. The ambiguity function in delay is widely used in radar systems to represent the distance resolution of the signal waveform. The sharper the ambiguity function, the better the distance resolution \cite{levanon2004radar}.  

Fig.\ref{fig:results_SP_a} shows the spectrum of the LFM signal with a bandwidth of 4 GHz at the TX.
Fig.\ref{fig:results_SP_b} shows the spectrum of the LFM signal at the DT after RIS beamforming, where the spectrum undergoes distortion and narrowband ``filtering" after RIS narrowband beamforming (the blue curve). On the contrary, the spectrum after the proposed RIS wideband beamforming maintains good fidelity with wideband characteristics (the red curve).
Moreover, Fig.\ref{fig:results_SP_c} shows the ambiguity function of the signals after RIS beamforming and the original LFM signal. Due to the distortion and narrowband ``filtering" caused by the RIS narrowband beamforming, the ambiguity function broadens, which causes worse distance resolution. On the contrary, the proposed wideband beamforming makes the gain large and flat over the desired bandwidth, thereby preserving the effective bandwidth as the signal bandwidth. Consequently, the ambiguity function keeps sharp after the proposed RIS wideband beamforming. Specifically, the distance resolution improves by more than two times compared with the narrowband beamforming. This suggests that the proposed SFFT-based RIS wideband beamforming offers a distinct advantage over RIS narrowband beamforming in sensing performance, particularly for applications that require high distance resolution, such as high-resolution imaging.

\section{Conclusions}
\label{sec:conclusion}
\textcolor{black}{In this paper, we propose a novel methodology by exploiting SFFT and SPM to solve the RIS wideband beamforming problem for the first time. 
Compared with previous RIS wideband beamforming methods, our method offers significant advantages. It does not require additional time-delay units and provides an approximate closed-form solution. This method allows a large and flat gain over the desired frequency band with extremely low complexity, which is adaptable to wideband applications with large RIS arrays. In particular, as the size of the RIS increases, the proposed beamforming shows improved flatness and a sharper roll-off. Moreover, simulation results demonstrate the advantages of the proposed wideband beamforming over the narrowband beamforming in terms of both communication rate and sensing resolution. Specifically, the communication rate improves by over 10\%. More importantly, we present a first-time finding that the distance resolution in sensing can be significantly enhanced with the RIS wideband beamforming. Through numerical results, it is shown that the distance resolution improves by more than two times compared with the narrowband beamforming.
The idea of applying SFFT for generating a flat beampattern is adaptable to generating any expected frequency-domain beampattern by matching the RIS phase shift with the amplitude modulation function. This provides valuable insights into the design of novel wideband beamforming for RIS-assisted systems.
}

\ifCLASSOPTIONcaptionsoff
  \newpage
\fi

\bibliographystyle{IEEEtran}
\bibliography{ref_bib}{}

\end{document}